\documentclass[aps, nofootinbib, pra, twocolumn]{revtex4-1}

\usepackage{ORI_Group_style}
\usepackage{epstopdf}
\usepackage{cleveref}

\newcommand{\CG}[2]{C_{#1}^{#2}}
\newcommand{\ue}[1]{\bold{e}_{#1}}

\newcommand{\Olab}{O\elx\ely\elz}
\newcommand{\elx}{\bold{e}_{\xl}}
\newcommand{\ely}{\bold{e}_{\yl}}
\newcommand{\elz}{\bold{e}_{\zl}}

\newcommand{\xl}{x}
\newcommand{\yl}{y}
\newcommand{\zl}{z}

\newcommand{\Obdy}{O\ebx\eby\ebz}
\newcommand{\ebx}{\bold{e}_{\xb}}
\newcommand{\eby}{\bold{e}_{\yb}}
\newcommand{\ebz}{\bold{e}_{\zb}}

\newcommand{\xb}{1}
\newcommand{\yb}{2}
\newcommand{\zb}{3}

\newcommand{\eula}{\alpha}
\newcommand{\eulb}{\beta}
\newcommand{\eulc}{\gamma}

\newcommand{\D}[2]{\mathcal{\hat{D}}_{#1}^{#2}}

\newcommand{\levi}[1]{\epsilon_{#1}}

\newcommand{\JJop}{\bold{\hat{J}}}

\newcommand{\Jop}{\hat{J}}
\newcommand{\Jlp}{\hat{J}_{+}}
\newcommand{\Jlm}{\hat{J}_{-}}
\newcommand{\Jlpm}{\hat{J}_{\pm}}

\newcommand{\LLop}{\bold{\hat{L}}}

\newcommand{\SSl}{\bold{\hat{F}}}
\newcommand{\SSb}{\bold{\hat{S}}}

\newcommand{\Ll}[1]{\hat{L}_{#1}}
\newcommand{\Lb}[1]{\hat{L}_{#1}}
\newcommand{\Llp}{\hat{L}_{+}}
\newcommand{\Llm}{\hat{L}_{-}}
\newcommand{\Llpm}{\hat{L}_{\pm}}

\newcommand{\Lbp}{\Lb{\uparrow}}
\newcommand{\Lbm}{\Lb{\downarrow}}

\newcommand{\Sl}[1]{\hat{F}_{#1}}
\newcommand{\Slp}{\hat{F}^{+}}
\newcommand{\Slm}{\hat{F}^{-}}

\newcommand{\Sb}[1]{\hat{S}_{#1}}
\newcommand{\Sbp}{\hat{S}_{\uparrow}}
\newcommand{\Sbm}{\hat{S}_{\downarrow}}

\newcommand{\Jl}[1]{\hat{J}_{#1}}
\newcommand{\Jb}[1]{\hat{J}_{#1}}
\newcommand{\Jbz}{\Jb{\zb}}
\newcommand{\Jbp}{\Jb{\uparrow}}
\newcommand{\Jbm}{\Jb{\downarrow}}

\newcommand{\Eang}{\hat{\Omega}}
\newcommand{\Jcn}{J}
\newcommand{\Lcn}{L}
\newcommand{\Scn}{S}
\newcommand{\Fcn}{F}
\newcommand{\M}[1]{m_{#1}}
\newcommand{\K}[1]{k_{#1}}

\newcommand{\EulRot}{R(\Eang)}
\newcommand{\ElemEulRot}[1]{R_{#1}(\Eang)}

\newcommand{\mom}{\boldsymbol{\hat{\mu}}}

\newcommand{\ppop}{\hat{\bold{p}}}

\newcommand{\UB}{\hat{U}}

\newcommand{\nnl}{\nn(\hat \rr)}
\newcommand{\nnb}{\nn(\hat \rr,\Eang)}
\newcommand{\nb}[1]{n_{#1}}

\newcommand{\nbz}{n_{\zb}(\Eang)}
\newcommand{\nbp}{n_{\uparrow}(\Eang)}
\newcommand{\nbm}{n_{\downarrow}(\Eang)}

\newcommand{\nl}[1]{n_{#1}}
\newcommand{\nlx}{n_{\xl}}
\newcommand{\nly}{n_{\yl}}
\newcommand{\nlz}{n_{\zl}}
\newcommand{\nlp}{n_{+}}
\newcommand{\nlm}{n_{-}}

\newcommand{\Vop}{\hat{V}}

\newcommand{\B}[1]{B''_{#1}}

\newcommand{\yop}{\hat{y}}

\newcommand{\LD}{\eta}
\newcommand{\dLD}{\eta'}

\newcommand{\TT}{T(\Eang)}

\newcommand{\zpm}[1]{r_{0#1}}

\newcommand{\wR}{\w_I}
\newcommand{\wB}{\w_L}
\newcommand{\wA}{\w_D}
\newcommand{\wT}[1]{\w_{#1}}
\newcommand{\wP}{\w_{T}}

\newcommand{\jop}{\hat{j}}
\newcommand{\jdop}{\hat{j}^\dag}
\newcommand{\mop}{\hat{m}}
\newcommand{\mdop}{\hat{m}^\dag}
\newcommand{\kop}{\hat{k}}
\newcommand{\kdop}{\hat{k}^\dag}
\newcommand{\sop}{\hat{s}}
\newcommand{\sdop}{\hat{s}^\dag}

\newcommand{\cop}[1]{\hat{b}_{#1}}
\newcommand{\cdop}[1]{\hat{b}_{#1}^{\dagger}}

\newcommand{\dop}{\hat{d}}
\newcommand{\ddop}{\hat{d}^{\dagger}}

\newcommand{\crop}{\hat{b}_{r}}
\newcommand{\crdop}{\hat{b}^{\dag}_{r}}
\newcommand{\clop}{\hat{b}_{l}}
\newcommand{\cldop}{\hat{b}^{\dag}_{l}}

\begin{document}

\title{Magnetic Rigid Rotor in the Quantum Regime: Theoretical Toolbox}

\author{Cosimo C. Rusconi}
\author{Oriol Romero-Isart}

\affiliation{Institute for Quantum Optics and Quantum Information of the
Austrian Academy of Sciences, A-6020 Innsbruck, Austria.}
\affiliation{Institute for Theoretical Physics, University of Innsbruck, A-6020 Innsbruck, Austria.}

\begin{abstract}
We describe the quantum dynamics of a magnetic rigid rotor in the mesoscopic scale where the Einstein-De Haas effect is predominant. In particular, we consider a single-domain magnetic nanoparticle with uniaxial anisotropy in a magnetic trap. Starting from the basic Hamiltonian of the system under the macrospin approximation, we derive a bosonized Hamiltonian describing the center-of-mass motion, the total angular momentum, and the macrospin degrees of freedom of the particle treated as a rigid body. This bosonized Hamiltonian can be approximated by a simple quadratic Hamiltonian that captures the rich physics of a nanomagnet tightly confined in position, nearly not spinning, and with its macrospin anti-aligned to the magnetic field. The theoretical tools derived and used here can be applied to other quantum mechanical rigid rotors. 
\end{abstract}

\maketitle

%%%%%%%%%%%%%%%%%%%%%%%%%%%%%%%%%%%%%%%%%%%%%%%%%%%%%%%%%%%%%%%%%
\section{Introduction}
%%%%%%%%%%%%%%%%%%%%%%%%%%%%%%%%%%%%%%%%%%%%%%%%%%%%%%%%%%%%%%%%%

A rigid body is described by its center-of-mass position and linear momentum as well as its orientation and rotational angular momentum~\citep{Landau:1960aa,Poole:1996aa}. A rigid body can also contain internal degrees of freedom. Particularly interesting is a magnetic rigid body such that the internal spin can couple to the external degrees of freedom via the Einstein-De Haas effect \citep{einstein1915experimental}. The physics of such a magnetic rigid rotor is very rich specially at mesoscopic scales where the Einstein-De Haas effect is enhanced due to a smaller moment of inertia. Quantum effects have to be considered~\citep{casimir1931rotation,WormerRigidRotor} when thermal fluctuations of the degrees of freedom are small. The quantum mechanical theory of the rigid rotor has been applied to study rotational spectra of molecules~\citep{RevModPhys.23.213,di2013rotational}, to model the structure of the atomic nucleus~\citep{bohr1969nuclear,RevModPhys.63.375}, and to control the rotational motion of molecules~\citep{PhysRevA.83.023423,PhysRevLett.112.113004,PhysRevLett.82.3420,arita2013laser}.
Magnetic rigid rotors in the quantum regime have been considered to study spin tunneling in single molecule magnets and magnetic single-domain nanoparticles~\citep{gatteschi2006molecular,O'Keeffe20122871,PhysRevB.81.214423,PhysRevLett.60.661,PhysRevLett.72.3433,ChudnoBook,PhysRevLett.104.027202}.

Motivated by the possibility to bring mesoscopic systems to the quantum regime, see \citep{RevModPhys.86.1391} and references therein, here we use well known results in quantum angular momentum theory~\citep{biedenharn1981angular,edmonds1996angular} to develop a theoretical toolbox to describe the quantum dynamics of a levitated magnetic rigid rotor.  We consider a single-domain magnetic nanoparticle (we call it nanomagnet hereafter) in a magnetic trap to derive, from first principles, a relatively simple quadratic Hamiltonian able to describe the rich dynamics of the system. A quantum mechanical description of the system has to be used when the thermal fluctuations of the degrees of freedom have been cooled to the limit where quantum fluctuations dominate. In the context of quantum nanomechanics many techniques have been devised to reach this regime \citep{RevModPhys.86.1391}. An experimental proposal to bring a levitated nanomagnet in the quantum regime will be analyzed elsewhere~\cite{CosimoInPreparation}.

The article is organized as follows.
In \secref{sec:System_Description} we introduce the system, its degrees of freedom, the basic Hamiltonian in the lab and body frame, and the Hilbert space with the relevant basis. In \secref{sec:Diag_MagneticDipole} we perform a unitary transformation that diagonalizes the magnetic dipole interaction term. 
In \secref{sec:Lamb-Dicke} we perform the Lamb-Dicke approximation by assuming the nanomagnet to be tighly confined in the magnetic trap.
In \secref{sec:Bosonization_AM} we show how to map the angular momentum operators of the Hamlitonian into bosonic creation and annihilation operators. This allows us to perform a Holstein-Primakov approximation by assuming the nanomagnet to be well anti-aligned to the external magnetic field and nearly not rotating. 
In \secref{sec:quadratic} we apply these approximations to derive a quadratic Hamiltonian and discuss its validity. 
In \secref{sec:Conclusions} we draw our conclusions and provide some further directions.
We leave to the appendices the discussion of some required tools of quantum angular momentum theory (\appref{apdx:Wigner_Dmatrix}),  an example of magnetic field trap  (\appref{apdx:Ioffe-Pritchard_Field}), and some  details about the derivation of the quadratic Hamiltonian (\appref{apdx:Calculation_V_D}  \& \ref{apdx:NonAdiabaticTerms}).

%%%%%%%%%%%%%%%%%%%%%%%%%%%%%%%%%%%%%%%%%%%%%%%%%%%%%%%%%%%%%%%%%
\section{Description of the System}\label{sec:System_Description}
%%%%%%%%%%%%%%%%%%%%%%%%%%%%%%%%%%%%%%%%%%%%%%%%%%%%%%%%%%%%%%%%%

\begin{figure*}[t]
	\includegraphics[width= 2\columnwidth]{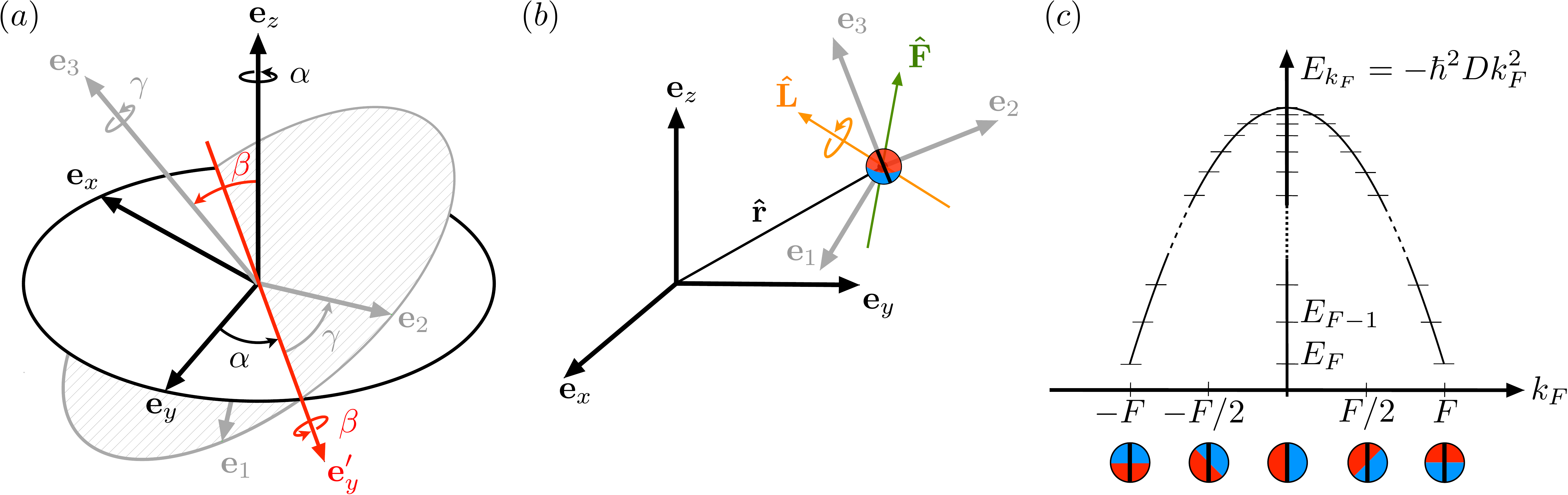}
	\caption{(a) Definition of the Euler angles in the ZYZ convention.  The Euler angles are defined by three successive rotations that one needs to apply in order to align $\Olab$ with $\Obdy$: (i) Rotation of the frame $\Olab$ of an angle $\eula\in [0, 2\pi]$ about the axis $\elz$ into the new frame $O\ue{x}'\ue{y}'\ue{z}'$. (ii) Rotation of the frame $O\ue{x}'\ue{y}'\ue{z}'$ of an angle $\eulb \in [0, \pi]$ about the axis $\ue{y}'$ into the new frame $O\ue{x}''\ue{y}''\ue{z}''$. (iii) Rotation of the new frame $O\ue{x}''\ue{y}''\ue{z}''$ of an angle $\eulc\in [0,2\pi]$ about the axis $\ue{z}''$ into the final frame $\Obdy$.
	(b) Description of the system and its degrees of freedom. The nanomagnet is represented as a sphere divided in two parts: a red and a blue half that stays respectively for the north and south pole of the magnet, and thus give the direction of the spin $\SSl$. The anisotropy axis is along the direction of $\ebz$ and it is represented by a black line in the nanomagnet which need not to be aligned with the magnetization. Finally the angular momentum $\LLop$ relative to the mechanical rotational motion lies, in general, along a different direction. 
	(c) Anisotropy of the nanomagnet. We consider a nanomagnet with a single axis anisotropy: the magnetic moment will preferably align parallel or antiparallel to the anisotropy direction. The $\Sl{\zb}^2$ dependence produce a non linear level structure.}
	\label{fig:Euler_Angles} 
\end{figure*}

We consider a nanomagnet of mass $M$, moment of inertia $I$, and magnetic moment $\mom$. The nanomagnet is interacting in free space with an external static B-field $\BB(\rr)$. Being a rigid body, the position of  the nanomagnet in space is specified by its center of mass position and its orientation. The latter is equivalent to specifying the orientation of the coordinate frame $\Obdy$, attached to the body and centered at the position of its center of mass, with respect to the position of the coordinate frame $\Olab$ fixed in the laboratory. 
The mutual orientation between the two frames is described through the Euler angles $\Omega=\{\eula,\eulb,\eulc\}$  defined in \figref{fig:Euler_Angles}a.
Hereafter Latin indexes $i,j,k,\ldots = \xb,\yb,\zb$ label the body frame axis while Greek indexes $\mu,\nu,\lambda\ldots =\xl,\yl,\zl$ label the laboratory frame axes. 

The nanomagnet is characterized by the following degrees of freedom, see Fig.~\ref{fig:Euler_Angles}b:
\begin{itemize}
	\item The center of mass described by its position $\hat{\rr}$ and momentum $\ppop$.
	\item The rotational momentum and orientation of the particle described by the angular momentum $\hbar\LLop$ and the Euler angles $\hat{\Omega}=\{\hat{\eula},\hat{\eulb},\hat{\eulc}\}$. 
	\item The magnetic moment $\mom$.
\end{itemize}

The operator $\hbar\LLop$ is the angular momentum associated to the rotational motion of the particle. The magnetic moment of the nanomagnet $\mom=\hbar\gamma \SSl$, with $ \SSl=\sum_{i=1}^N \SSl_i $,  is the sum of the single magnetic moments of its $N$ constituents, where $\gamma>0$ is the gyromagnetic ratio and $\SSl_i$ the total spin of the $i$th constituent.
The individual spins inside the nanomagnet are subjected to two different interactions: (i) the exchange interaction, that tends to align all the single magnetic moments together, and (ii) the anisotropy interaction, that tends to align independently each magnetic moment to a given direction in the crystalline structure of the particle.  When the first interaction dominates over the second (weak anisotropy limit), one can perform the so-called macrospin approximation~ \citep{sayad2012macrospin}, which consists in projecting the total spin into the subspace with $\SSl^2 = N f(N f+1) \equiv F(F+1)$, where $f$ is the total spin of a single constituent (assumed to be identical for simplificity). Finally, note that the center-of-mass motion, described by $\hat{\rr}$ and $\ppop$, can yield angular momentum $\hat{\rr} \times \ppop$. This should not be confused with $\hbar\LLop$.

For a solid nanosphere the internal vibrations have frequencies of the order of the speed of sound divided by the size of the object. For a nanosphere, these frequencies are many orders of magnitude higher  than any other frequency in the system. Hence, internal vibrations are effectively decoupled~\citep{PhysRevA.83.013803} and can be safely ignored.

\subsection{Hamiltonian in the Laboratory Frame}\label{sec:Sys_Ham_Lab}

The dynamics of a rigid body can be studied both in the laboratory frame $\Olab$ and in the body frame $\Obdy$. 
In the laboratory frame, the dynamics of a spherical nanomagnet of radius $R$ (moment of inertia $I=2MR^2/5$) is described by the Hamiltonian
\be\label{eq:Ham_Lab}
	\Hop = \frac{\ppop^2}{2M}+\frac{\hbar^2}{2I}\LLop^2 -\hbar^2D \spare{\SSl\cdot\ebz(\Eang)}^2 -\hbar\gamma\SSl\cdot\BB(\hat \rr).
\ee
Note that when a non-spherical shape is considered, the moment of inertia will be a tensor depending on the Euler angles \citep{Landau:1960aa,Poole:1996aa}.
The first (second) term represents the kinetic energy of the center of mass (rotational) motion. 
The third term represents the uniaxial anisotropy energy term in the macrospin approximation with the preferred axis $\ebz$, see \figref{fig:Euler_Angles}c. The uniaxial anisotropy is common in nanomagnets \citep{gatteschi2006molecular} but other anisotropies could be straightforwardly included in the analysis.
In the Laboratory frame, the unit vector $\ebz$ depends on the Euler angles and it is therefore an operator, $\ebz=\ebz(\Eang)$.
The anisotropy constant $D$ depends on the material and is related to the blocking temperature of the nanomagnet $T_b \equiv \hbar^2 DF^2/K_b$ \citep{bahmanrokh2013simple}, where $K_b$ is the Boltzmann's constant.
The fourth term represents the magnetic dipole interaction of the macrospin with the external static field $\BB(\rr)$ which we describe as a classical field.

For later convenience we define $\JJop \equiv \LLop + \SSl$, which is the total angular momentum of the system when $\hat \rr \times \ppop =0$. The operators $\JJop,\LLop,$ and $\SSl$ fulfill the usual commutation relations of angular momenta, see \tabref{TAB:commut}. Moreover, we assume $[\Ll{\nu},\Sl{\mu}]=0, \,\forall \mu,\nu$, namely that the spin angular momentum is independent of the Euler angles. This is commonly assumed in molecular quantum mechanics \citep{brown2003rotational,RevModPhys.23.213}, and corresponds to neglecting the spin-orbit interaction between the individual spins of the nanomagnet and the electronic rotational motion within the Born-Oppenheimer approximation.

\begin{table*}[t]
\caption{Commutation rules of the components of the angular momentum operators $\LLop$, $\SSb=-\SSl$ and $\JJop = \SSl+\LLop = -\SSb + \LLop$ both in the laboratory frame (left column) and in the body fixed frame (right column). The ladder operators in the body frame  are $\Jb{\uparrow(\downarrow)} = \Jb{1} \mp \im \Jb{2} $ ($\Sb{\uparrow(\downarrow)} = \Sb{1} \mp \im \Sb{2}$), while in the laboratory are $\Jl{\pm}=\Jl{\xl}\pm \im \Jl{\yl}$ ($\Sl{\pm}=\Sl{\xl}\pm \im \Sl{\yl}$). Summation over repeated indexes is here assumed.}\label{TAB:commut}
\begin{ruledtabular}
\begin{tabular}{|cc|}
Laboratory Frame $\Olab$ & Body Frame $\Obdy$\\
\hline
&\\
$\begin{array}{rcl}
\com{\Ll{\mu}}{\Ll{\nu}}&=&\im\levi{\mu\nu\lambda}\Ll{\lambda}\\
\com{\Ll{\zl}}{\Llpm}&=&\pm\Llpm\\
\com{\Llp}{\Llm}&=& 2\Ll{\zl}\\
\com{\Sl{\mu}}{\Sl{\nu}}&=&\im\levi{\mu\nu\lambda}\Sl{\lambda}\\
\com{\Sl{\zl}}{\Sl{\pm}}&=&\pm\Sl{\pm}\\
\com{\Slp}{\Slm}&=&2\Sl{\zl}\\
\com{\Jl{\mu}}{\Jl{\nu}}&=&\im\levi{\mu\nu\lambda}\Jl{\lambda}\\
\com{\Jl{\zl}}{\Jlpm}&=&\pm\Jlpm\\
\com{\Jlp}{\Jlm}&=& 2\Jl{\zl}\\
\com{\Ll{\mu}}{\Sl{\nu}}&=&0\\
\com{\Jl{\mu}}{\Sl{\nu}}&=&\im\levi{\mu\nu\lambda}\Sl{\lambda}\\
\com{\Jl{\mu}}{\Ll{\nu}}&=&\im \levi{\mu\nu\lambda}\Ll{\lambda}\\
\com{\Ll{\mu}}{\ElemEulRot{i\nu}}&=&\im\levi{\mu\nu\lambda}\ElemEulRot{i\lambda}\quad \forall i\\
\com{\Jl{\mu}}{\ElemEulRot{i\nu}}&=&\im\levi{\mu\nu\lambda}\ElemEulRot{i\lambda}\quad \forall i\\
\com{\Sl{\mu}}{\ElemEulRot{i\nu}}&=&0\\
\com{\ElemEulRot{i\mu}}{\ElemEulRot{j\nu}}&=&0\quad \forall i,j,\mu,\nu\\
\end{array}$
&
$\begin{array}{rcl}
\com{\Lb{i}}{\Lb{j}}&=&-\im\levi{ijk}\Lb{k}\\
\com{\Lb{\zb}}{\Lb{\uparrow(\downarrow)}}&=&(-)\Lb{\uparrow(\downarrow)}\\
\com{\Lbp}{\Lbm}&=& 2\Lb{\zb}\\
\com{\Sb{i}}{\Sb{j}}&=&-\im\levi{ijk}\Sb{k}\\
\com{\Sb{\zb}}{\Sb{\uparrow(\downarrow)}}&=&(-)\Sb{\uparrow(\downarrow)}\\
\com{\Sbp}{\Sbm}&=& 2\Sb{\zb}\\
\com{\Jb{i}}{\Jb{j}}&=&-\im\levi{ijk}\Jb{k}\\
\com{\Jb{\zb}}{\Jb{\uparrow(\downarrow)}}&=&(-)\Jb{\uparrow(\downarrow)}\\
\com{\Jbp}{\Jbm}&=& 2\Jb{\zb}\\
\com{\Lb{i}}{\Sb{j}}&=&-\im\levi{ijk}\Sb{k}\\
\com{\Jb{i}}{\Sb{j}}&=&0\\
\com{\Jb{i}}{\Lb{j}}&=&-\im \levi{ijk}\Jb{k}\\
\com{\Lb{i}}{\ElemEulRot{j\mu}}&=&-\im\levi{ijk}\ElemEulRot{k\mu}\quad \forall \mu\\
\com{\Jb{i}}{\ElemEulRot{j\mu}}&=&-\im\levi{ijk}\ElemEulRot{k\mu}\quad \forall \mu\\
\com{\Sb{i}}{\ElemEulRot{j\mu}}&=&0\\
\com{\ElemEulRot{i\mu}}{\ElemEulRot{j\nu}}&=&0\quad \forall i,j,\mu,\nu\\
\end{array}$
\\
&\\
\end{tabular}
\end{ruledtabular}
\end{table*}

\subsection{Hamiltonian in the Body Frame}\label{sec:BF_Hamiltonian}

The dynamics of a rigid body is more conveniently described in the body frame $\Obdy$ where the inertia tensor of a non-spherical body does not depend on the Euler angles \citep{Landau:1960aa,Poole:1996aa}. The operators in the body frame are obtained by the following change of variables:
\be\label{eq:Lab2Body}
\begin{split}
	&\Jb{i}=\sum_\nu \ElemEulRot{i\nu}\Jl{\nu},\\
	&\Ll{i}=\sum_\nu \ElemEulRot{i\nu}\Ll{\nu},\\
	&\Sl{i}=\sum_\nu \ElemEulRot{i\nu}\Sl{\nu},\\
	&B_i(\rr,\Eang) = \sum_\nu \ElemEulRot{i\nu}B_\nu(\rr),\\
\end{split}
\ee
where the orthogonal matrix $R(\Eang)$ is given by
\be\label{eq:EulRot}
\begin{split}
	R(\Eang)=&
		\begin{pmatrix}
			\cos\hat{\eulc} & \sin\hat{\eulc} & 0\\
			-\sin\hat{\eulc} & \cos\hat{\eulc} & 0\\
			0 & 0 & 1
		\end{pmatrix}
		\begin{pmatrix}
			\cos\hat{\eulb} & 0 & -\sin\hat{\eulb}\\
			0 & 1 & 0\\
			\sin\hat{\eulb} & 0 & \cos\hat{\eulb}
		\end{pmatrix}\\
		&\times \begin{pmatrix}
			\cos\hat{\eula} & \sin\hat{\eula} & 0\\
			-\sin\hat{\eula} & \cos\hat{\eula} & 0\\
			0 & 0 & 1
		\end{pmatrix}.
\end{split}
\ee

The change of variables \eqnref{eq:Lab2Body} does not preserve the commutation relations of the angular momenta. 
Indeed, by writing the operator $\LLop$ in the laboratory frame in the Euler angles representation~\citep{edmonds1996angular}
\be\label{eq:Llab}
	\vect{\Ll{\xl}}{\Ll{\yl}}{\Ll{\zl}}= \frac{\im}{\sin\eulb}
	\begin{pmatrix}
			\cos\eula\cos\eulb & \sin\eulb\sin\eula &-\cos\eula\\
			\cos\eulb\sin\eula & -\sin\eulb\cos\eula & -\sin\eula\\
			-\sin\eulb & 0 & 0
	\end{pmatrix}
	\vect{\pa{\eula}}{\pa{\eulb}}{\pa{\eulc}},
\ee
one can show that using \eqnref{eq:EulRot}
\be\label{eq:commutLlabR}
	\com{\Ll{\mu}}{\ElemEulRot{i\nu}}=\im \sum_{\lambda}\levi{\mu\nu\lambda}\ElemEulRot{i\lambda}.
\ee
Note however that this is not the case for the spin angular momentum since, as already discussed in \secref{sec:Sys_Ham_Lab}, it commutes with the Euler angles 
\be\label{eq:commutSlabR}
	\com{\Sl{\mu}}{\ElemEulRot{i\nu}}=0.
\ee
The results in \eqnref{eq:commutLlabR} and in \eqnref{eq:commutSlabR} together with the property
\be
   \sum_{\mu,\nu} \levi{\mu\nu\lambda}\ElemEulRot{i\mu}\ElemEulRot{j\nu} = \sum_k \levi{ijk}\ElemEulRot{k\lambda},
\ee
valid for an orthogonal matrix, allow to derive the commutation relations of the operators in the body frame, as given in \tabref{TAB:commut}. Note that the body frame components of the spin commute as in the laboratory frame, while the commutators for the body frame components of $\LLop$ and $\JJop$ acquire a minus sign, namely $[\Lb{i},\Lb{j}]=-\im \sum_k \levi{ijk}\Lb{k}$ and $[\Jb{i},\Jb{j}]=-\im \sum_k \levi{ijk} \Jb{k}$. 
It is convenient to introduce the operator $\SSb = -\SSl$ to force the three angular momenta to have the same commutation rules. Note that $\SSb^2 = \SSl^2$ and we define $\Scn \equiv \Fcn$. The body frame ladder operators are given by $\Jbp= (\Jbm)^{\dag} = \Jb{\xb} - \im \Jb{\yb}$ and $\Sbp= (\Sbm)^\dag = \Sb{\xb} - \im \Sb{\yb}$, where $\Jbp, \Sbp$ ($\Jbm,\Sbm$) increase (lower) the value of $\Jb{\zb}$, $\Sb{\zb}$, see \tabref{TAB:commut}.

With the change of variables given in \eqnref{eq:Lab2Body}, the Hamiltonian $\Hop$ in the body frame reads
\be\label{eq:Ham_Body}
\begin{split}
	\Hop =& \frac{\ppop^2}{2M} + \frac{\hbar^2}{2I}\pare{\JJop^2 +2\Sb{\zb}\Jb{\zb}+\Jbp\Sbm+\Jbm\Sbp}\\
	& -\hbar^2 D\Sb{\zb}^2 +\hbar\gamma \SSb\cdot\BB(\hat \rr,\Eang).
\end{split}
\ee
The term with $\SSb^2$ has been dropped since under the macroscopic approximation $\SSb^2=S(S+1)$ is a $c$-number. The external magnetic field depends in this frame on the Euler angles. So far only a change of variables has been performed without any approximation. 

\subsection{Hilbert Space Structure of the System and Suitable Basis}\label{sec:HilbertSpace_Struct}

The Hilbert space $\mathcal{H}$ associated to the system is given by
	$\mathcal{H} = \mathbb{L}^2\pare{\mathbb{R}^3}\otimes \mathcal{H}_{am}$.
$\mathbb{L}^2\pare{\mathbb{R}^3}$ is the Hilbert space associated to the motion of a point-like particle in $\mathbb{R}^3$, and it is spanned by the general basis $\{\psi_n(\rr)\}_n$. $\mathcal{H}_{am}$ is the Hilbert space associated to the angular momenta of the particle. From the commutation relations in \tabref{TAB:commut} it is possible to find three different complete sets of commuting observables (CSCO) for the particle angular momenta~\citep{di2013rotational}:
\begin{enumerate}
	\item Laboratory frame uncoupled representation $\{\SSl^2,\Sl{\zl},\LLop^2,\Ll{\zl},\Lb{\zb}\}$ with eigenstates
	\be\label{eq:KetLab}
		\ket{\varphi_{am}}_1=\ket{\Fcn \M{\Fcn},\Lcn \M{\Lcn} \K{\Lcn}}_1,
	\ee
	such that
	\be
		\begin{pmatrix}
		\SSl^2 \\
		\Sl{\zl}\\
		\LLop^2\\
		\Ll{\zl}\\
		\Ll{\zb}
		\end{pmatrix} \ket{\varphi_{am}}_1 = 
		\begin{pmatrix}
		\Fcn(\Fcn+1) \\
		\M{\Fcn}\\
		\Lcn(\Lcn+1)\\
		\M{\Lcn}\\
		\K{\Lcn}.
		\end{pmatrix}\ket{\varphi_{am}}_1.
	\ee	
	The Hilbert space is $\mathcal{H}_{am}^1=\mathbb{C}^{2\Fcn+1}\otimes\big(\bigoplus_{L=0}^{\infty} \mathbb{C}^{(2\Lcn+1)^2}\big)$.
	\item Body frame uncoupled representation $\{\JJop^2,\Jl{\zl},\Jb{\zb},\SSb^2,\Sb{\zb}\}$ with eigenstates
	\be\label{eq:KetBody}
		\ket{\varphi_{am}}_2=\ket{\Jcn \M{\Jcn} \K{\Jcn}, \Scn \K{\Scn}}_2,
	\ee
	such that
	\be
		\begin{pmatrix}
		\SSb^2 \\
		\Sb{\zb}\\
		\JJop^2\\
		\Jl{\zl}\\
		\Jb{\zb}
		\end{pmatrix} \ket{\varphi_{am}}_2 = 
		\begin{pmatrix}
		\Scn(\Scn+1) \\
		\M{\Scn}\\
		\Jcn(\Jcn+1)\\
		\M{\Jcn}\\
		\K{\Jcn}
		\end{pmatrix}\ket{\varphi_{am}}_2.
	\ee	
	The Hilbert space is $\mathcal{H}_{am}^2=\mathbb{C}^{2\Scn+1}\otimes\big(\bigoplus_{J=0}^{\infty}\mathbb{C}^{(2\Jcn+1)^2}\big)$.
	\item Coupled representation $\{\JJop^2,\Jl{\zl},\SSb^2,\LLop^2,\Lb{\zb}\}$ with eigenstates
	\be	\label{eq:KetCoupled}
		\ket{\varphi_{am}}_3=\ket{\Jcn \M{\Jcn},\Scn,\Lcn \K{\Lcn}}_3,
	\ee
	such that
	\be
		\begin{pmatrix}
		\JJop^2\\
		\Jl{\zl}\\
		\SSb^2 \\
		\LLop^2\\
		\Lb{\zb}
		\end{pmatrix} \ket{\varphi_{am}}_3 = 
		\begin{pmatrix}
		\Jcn(\Jcn+1) \\
		\M{\Jcn}\\
		\Scn(\Scn+1)\\
		\Lcn(\Lcn+1)\\
		\K{\Lcn}
		\end{pmatrix}\ket{\varphi_{am}}_3.
	\ee	
	The Hilbert space is $\mathcal{H}_{am}^3 =\bigoplus_{L=0}^{\infty}\pare{\mathbb{C}^{2\Lcn+1}\otimes\mathbb{C}^{d_j}} \cong \bigoplus_{J=0}^{\infty}\pare{\mathbb{C}^{2\Jcn+1}\otimes\mathbb{C}^{d_l}}$, where $d_j = \sum_{\Jcn=|\Lcn-\Fcn|}^{\Lcn+\Fcn}2\Jcn+1$ and $d_l = \sum_{L=|\Jcn-\Scn|}^{\Jcn+\Scn}2\Lcn+1$.
\end{enumerate}

The set $1$ ($2$) contains the additional commuting operator $\Jl{\zl}=\Sl{\zl}+\Ll{\zl}$ ($\Lb{\zb}=\Sb{\zb}+\Jl{\zb}$), which has not been included since it is determined by some of the operators within the set and thus it does not represent an additional degree of freedom. 
The three Hilbert spaces of the representation are isomorphic (they correspond to a different choice of basis), and the change of basis is achieved by a unitary transformation. To switch from $1$ to $3$ one uses the relation
\be \label{eq:H1toH2}
\ket{\Fcn\M{\Fcn},\Lcn\M{\Lcn}\K{\Lcn}}_1= \! \! \! \! \! \! \sum_{\Jcn=|\Lcn-\Fcn|}^{\Lcn+\Fcn} \! \! \!  \CG{\Fcn\M{\Fcn},\Lcn\M{\Lcn}}{\Jcn\M{\Jcn}}\ket{\Jcn\M{\Jcn},\Scn,\Lcn\K{\Lcn}}_3
\ee
and from $2$ to $3$ the relation
\be  \label{eq:H2toH3}
\ket{\Jcn\M{\Jcn}\K{\Jcn},\Scn \K{\Scn}}_2= \! \! \! \! \! \! \sum_{\Lcn=|\Jcn-\Scn|}^{\Jcn+\Scn} \! \! \!   \CG{\Jcn\K{\Jcn},\Scn \K{\Scn}}{\Lcn\K{\Lcn}}\ket{\Jcn\M{\Jcn},\Scn,\Lcn\K{\Lcn}}_3.
\ee
Here $\CG{\Fcn\M{\Fcn},\Lcn\M{\Lcn}}{\Jcn\M{\Jcn}}={}_3\braket{\Jcn\M{\Jcn},\Scn,\Lcn\K{\Lcn}}{\Fcn\M{\Fcn},\Lcn\M{\Lcn}\K{\Lcn}}_{1}$ and $\CG{\Jcn\K{\Jcn},\Scn \K{\Scn}}{\Lcn\K{\Lcn}}={}_3\braket{\Jcn\M{\Jcn},\Scn,\Lcn\K{\Lcn}}{\Jcn\M{\Jcn}\K{\Jcn},\Scn \K{\Scn}}_2$ are Clebsch-Gordan coefficients under the Condon-Shortley convention (they are real numbers)~\citep{edmonds1996angular}. $\CG{J_1m_1,J_2m_2}{Jm}$ is non-zero provided $m=m_1+m_2$ and $J\in \{J_1+J_2,J_1+J_2-1,\ldots,|J_1-J_2|\}$, which embodies the conservation of angular momentum.

The inspection of the Hamiltonian in the body frame, \eqnref{eq:Ham_Body}, suggests to use the representation $2$ (body frame uncoupled representation). The basis of the total Hilbert space will thus be $\{\ket{\psi_n}\otimes\ket{\Jcn \M{\Jcn} \K{\Jcn}, \Scn \K{\Scn}}_2\}$.

%%%%%%%%%%%%%%%%%%%%%%%%%%%%%%%%%%%%%%%%%%%%%%%%%%%%%%%%%%%%%%%%%
\section{Diagonalization of the Magnetic Dipole Interaction}\label{sec:Diag_MagneticDipole}
%%%%%%%%%%%%%%%%%%%%%%%%%%%%%%%%%%%%%%%%%%%%%%%%%%%%%%%%%%%%%%%%%

\begin{figure}
	\includegraphics[width= 0.5 \columnwidth]{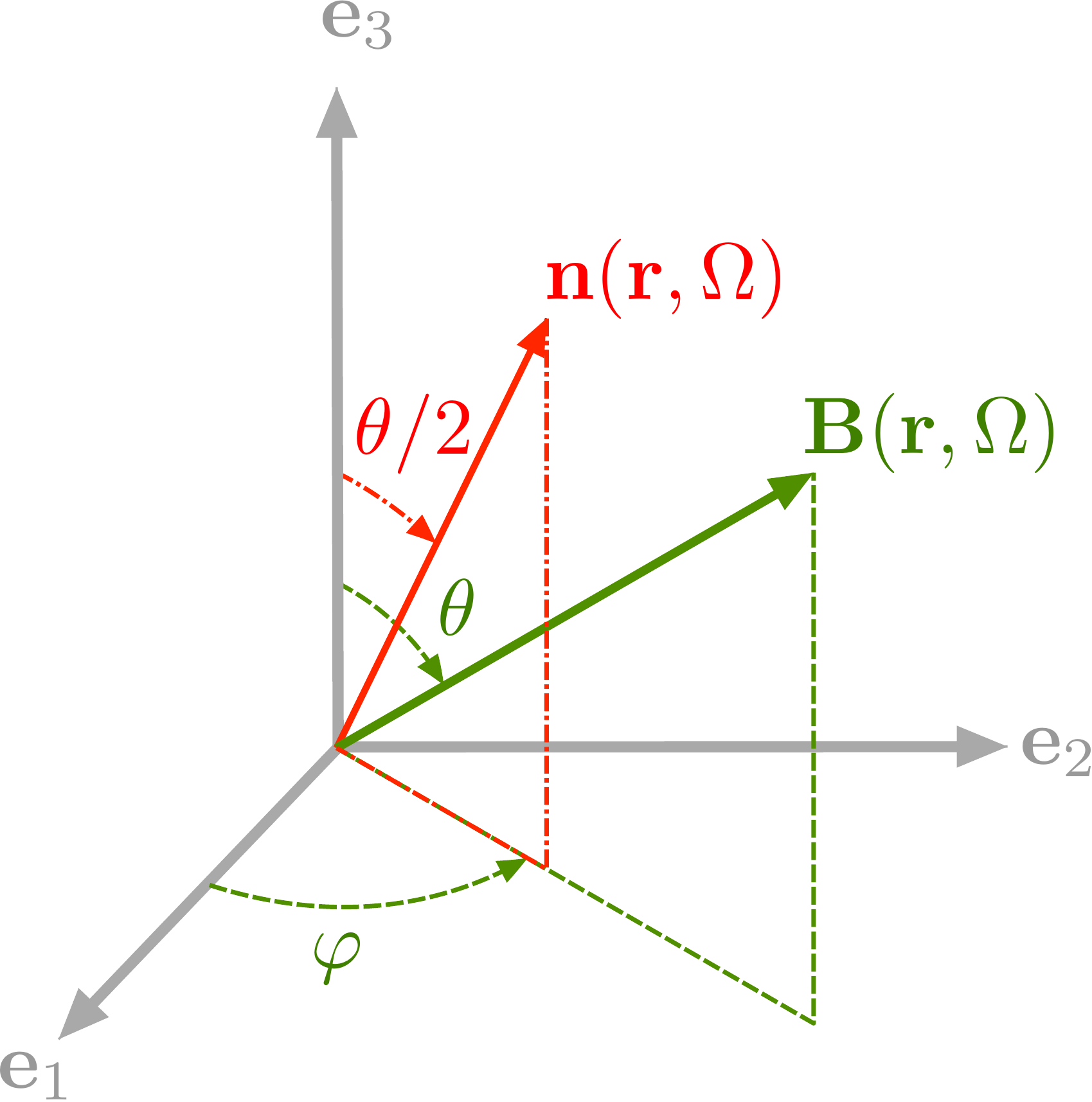}
	\caption{Definition of the vector $\nn(\rr,\Omega)$ and $\BB(\rr,\Omega)$. The unitary $\UB$ is a rotation of an angle $\pi$ around the local direction $\nn(\rr,\Omega)$. This transformation aligns $\ebz$ with $\BB(\rr,\Omega)$.}
	\label{fig:U_rotation_scheme}
\end{figure}

In the context of magnetic trapping it is convenient to apply a unitary transformation  $\UB$ to the Hamiltonian such that the magnetic dipole interaction in \eqnref{eq:Ham_Body} is transformed as follows:
\be
	\UB^\dag \BB(\hat \rr,\Eang)\cdot \SSb \UB = \vert\BB(\hat \rr)\vert \Sb{\zb}.
\ee
This can be achieved with the unitary transformation given by\footnote{An analogous unitary transformation is also used in describing the magnetic trapping of atoms~\citep{PhysRevA.56.2451,PhysRevA.74.035401}. However, note that here the unitary operator \eqnref{eq:Unitary_Body} depends on $\Eang$ and hence has to be treated more carefully.}
\be\label{eq:Unitary_Body}
	\UB \equiv \exp\spare{-\im\pi \nnb\cdot\SSb},
\ee
where $\nnb=(\nb{\xb}(\hat \rr,\Eang),\nb{\yb}(\hat \rr,\Eang),\nb{\zb}(\hat \rr,\Eang))^T$ is the unit vector that lies in the same plane as $\ebz$ and $\BB(\hat \rr,\Eang)$, and bisects the angle between them, see \figref{fig:U_rotation_scheme}.

The body frame Hamiltonian \eqnref{eq:Ham_Body} transformed by the unitary  transformation \eqnref{eq:Unitary_Body} can be written as
\be\label{eq:Unitary_Ham_def}
	\Hop' = \UB^{\dag}\Hop\UB = \Hop_0 + \Hop_I + \Vop_D + \Vop_P,
\ee
where:
\begin{itemize}
\item $\Hop_0$, which is diagonal in the basis $\ket{\Jcn \M{\Jcn} \K{\Jcn}, \Scn \K{\Scn}}_2$, is given by
\be\label{eq:H_0}
\begin{split}
	\Hop_0 =& \frac{\ppop^2}{2M} + \frac{\hbar^2}{2I}\pare{\JJop^2+2\Sb{\zb}\Jb{\zb}} -\hbar^2D\Sb{\zb}^2 \\& +\hbar\gamma|\BB(\hat \rr)|\Sb{\zb}  	.
\end{split}
\ee
Since local minima of $|\BB(\hat \rr)|$ are allowed, the term $\hbar\gamma|\BB(\hat \rr)|\Sb{\zb}$ can trap the nanomagnet for macrospin states with $\avg{\Sb{\zb}} >0 $, .

\item $\Hop_I$ is given by
\be\label{eq:V_I}
		\Hop_{I} =\frac{\hbar^2}{2I}\pare{\Jb{\uparrow}\Sb{\downarrow}+\Jb{\downarrow}\Sb{\uparrow}}.
\ee
This term originates from the rotational kinetic energy of the nanomagnet, $\hbar(\JJop+\SSb)^2/2 I$ and couples the spin $\SSb$ with the total angular momentum $\JJop$.

\item $\Vop_D$ is given by
\be\label{eq:V_D}
\begin{split}
	 \Vop_D =& \hbar^2D\Sb{\zb}^2 -\hbar^2D\bigg[\Sb{\zb}\pare{1-2\nb{\zb}^2(\hat \rr,\Eang)}\\
	&-\nb{\zb}(\hat \rr,\Eang)\pare{\nb{\uparrow}(\hat \rr,\Eang)\Sbm+ \nb{\downarrow}(\hat \rr,\Eang)\Sbp}\bigg]^2,
\end{split}
\ee
where we defined $\nb{\uparrow}(\hat \rr,\Eang)= [\nb{\downarrow}(\hat \rr,\Eang)]^\dag  = \nb{\xb}(\hat \rr,\Eang)- \im \nb{\yb}(\hat \rr,\Eang)$. This arises from the unitary transformation of the anisotropy interaction.

\item $\Vop_P$ is given by
\be\label{eq:V_P}
\begin{split}
\Vop_P =& \frac{\ppop\cdot \AB(\hat \rr,\Eang) +\AB(\hat \rr,\Eang)\cdot\ppop}{2M} +\frac{\AB^2(\hat \rr,\Eang)}{2M},
\end{split}
\ee
 where $A_{\nu}(\hat \rr,\Eang)=-2\hbar\spare{\nnb\times\pa{\nu}\nnb}\cdot\SSb$. This term originates in the unitary transformation of the center of mass momentum $\ppop$, which does not commute with $\UB$.
\end{itemize}

The transformed Hamiltonian \eqnref{eq:Unitary_Ham_def} is still exact. The terms $\Vop_D$ and $\Vop_P$ contain the components of $\nnb$, which are operators acting both on the center of mass subspace, through $\hat \rr$, and on the angular momentum subspace, through the Euler angles $\Eang$. One can show that
\be\label{eq:Connerion_nb_nl}
	\nnb = \EulRot\, \nnl,
\ee
where  $\nn(\hat \rr)$ is the unit vector that lies in the same plane as $\BB(\hat \rr)$ and $\elz$ and bisects the angle between the two. The operator $\nnb$ acts on the angular momentum subspace through the matrix elements $\ElemEulRot{i\nu}$, which are combination of trigonometric functions of the Euler angles, see \eqnref{eq:EulRot}. These can be expressed through the Wigner D-matrix tensor operators $\D{mk}{j}$ \citep{Sakurai}, which are defined as follows. Consider the function $D_{mk}^{j}(\eula,\eulb,\eulc)$ as the matrix element of the general unitary rotation operator in quantum mechanics\footnote{The ket $\ket{jm}$ is defined as $\JJop^2 \ket{jm} = j(j+1) \ket{jm}$ and $\Jop_z \ket{jm} = m \ket{jm}$, where $\JJop$ is here a generic angular momentum operator.}
\be\label{eq:WignerD}
	D_{mk}^{j}(\eula,\eulb,\eulc) \equiv \bra{jk}e^{\im \eulc \Jop_z}e^{\im \eulb \Jop_y}e^{\im \eula\Jop_z} \ket{jm},
\ee
where $e^{\im \eulc \Jop_z}e^{\im \eulb \Jop_y}e^{\im \eula \Jop_z}$ is the unitary representation of the Euler angles rotation matrix according to the convention adopted in \figref{fig:Euler_Angles}. In \eqnref{eq:WignerD} the angles $\eula,\eulb,\eulc$ are real parameters. The Wigner D-matrix tensor operator is then defined by $\D{mk}{j}\equiv D_{mk}^{j}(\hat{\eula},\hat{\eulb},\hat{\eulc})$, that is by promoting the Euler angles to operators. Hereafter we call the D-matrix tensor operator just D-matrix.
All the elements of the rotation matrix $\EulRot$ can be written as linear combination of $\D{mk}{j}$. The rotation matrix of \eqnref{eq:EulRot} reads
\begin{widetext}
\be\label{eq:EulRot_Dmatrix}
R(\Eang)=\frac{1}{2}
	\begin{pmatrix}
		\pare{\D{11}{1}+\D{-1-1}{1}-\D{1-1}{1}-\D{-11}{1}} & -\im\pare{\D{11}{1}-\D{-1-1}{1}+\D{-11}{1}-\D{1-1}{1}} & \sqrt{2}\pare{\D{0-1}{1}-\D{01}{1}}\\
		\im\pare{\D{11}{1}-\D{-1-1}{1}-\D{-11}{1}+\D{1-1}{1}}& \pare{\D{11}{1}+\D{-1-1}{1}+\D{1-1}{1}+\D{-11}{1}} & -\im\sqrt{2}\pare{\D{0-1}{1}+\D{01}{1}}\\
		\sqrt{2}\pare{\D{-10}{1}-\D{10}{1}} & \im\sqrt{2}\pare{\D{10}{1}+\D{-10}{1}} & 2\D{00}{1}
	\end{pmatrix}.
\ee
\end{widetext}
This allows to write $\Vop_D$ and $\Vop_P$ in terms of D-matrices, since using \eqnref{eq:Connerion_nb_nl} one has that 
\be\label{eq:nnl2nnb}
	\vect{\nb{\zb}(\hat{\rr},\Eang)}{\nb{\uparrow}(\hat{\rr},\Eang)}{\nb{\downarrow}(\hat{\rr},\Eang) }= \TT \vect{\nl{\zl}(\hat{\rr})}{\nlp(\hat{\rr})}{\nlm(\hat{\rr})},
\ee
where $\nl\pm(\hat{\rr}) \equiv \nlx(\hat{\rr})\pm\im \nly(\hat{\rr})$ and 
\be \label{eq:T_matrix} \\
	\TT \equiv \frac{1}{\sqrt{2}}
		\begin{pmatrix}
			\sqrt{2}\D{00}{1} & \D{-10}{1} & -\D{10}{1}\\
			-2\D{01}{1} & -\sqrt{2}\D{-11}{1} & \sqrt{2}\D{11}{1}\\
			2\D{0-1}{1} & \sqrt{2}\D{-1-1}{1} & -\sqrt{2}\D{1-1}{1}
		\end{pmatrix}.
\ee
As shown in \appref{apdx:Wigner_Dmatrix},  the D-matrices transform the state $\ket{\Jcn \M{\Jcn} \K{\Jcn}, \Scn \K{\Scn}}_2$ in the following way
\be\label{eq:Action_Dmatrix}
\begin{split}
	&\D{mk}{j} \ket{\Jcn \M{\Jcn} \K{\Jcn}, \Scn \K{\Scn}}_2= \\
	&= \sum_{J'=|\Jcn-j|}^{\Jcn+j} \sqrt{\frac{2\Jcn +1}{2J'+1}}\CG{jm,J\M{J}}{J'\M{J}'}\CG{jk,J\K{J}}{J'\K{J}'}\,\ket{J'\M{\Jcn}'\K{\Jcn}',  \Scn \K{\Scn}}_2,
\end{split}
\ee
where $\M{\Jcn}'=\M{\Jcn}+m$ and $\K{\Jcn}'=\K{\Jcn}+k$. For $j \neq 0$, the D-matrix operators couple subspaces of different $J$.

The Hamiltonian $\Hop'$, \eqnref{eq:Unitary_Ham_def}, which is derived from $\Hop$, \eqnref{eq:Ham_Body}, without any approximation, can be expressed as a function of the following operators: $\hat \rr$, $\hat \pp$, $ \Sb{3}$, $\Sb{\uparrow(\downarrow)}$, $\JJop^2$, $\Jop_{3}$, $\Jop_{z}$, $ \Jop_{\uparrow(\downarrow)}$, and products of $\D{mk}{1}$ (for $m,k=-1,0,1$)\footnote{As shown in \appref{apdx:Wigner_Dmatrix}, products of D-matrices can always be written as a linear combination of D-matrices $\D{mk}{j}$ with $j\geq 1$.}. That is, it can be expressed without an explicit dependence on the Euler angles. We remark that in such a transformed frame, some of the mathematical operators appearing in $\Hop'$, \eqnref{eq:Unitary_Ham_def}, have a different physical meaning that those appearing in $\Hop$, \eqnref{eq:Ham_Body}. In particular note that (we define $\hat M' \equiv \UB^\dag  \hat M \UB  $ for any operator $\hat M$),
\bea
    \hat{\rr}' &=& \hat{\rr}, \\
    \ppop' &=& \ppop + \AB(\hat \rr,\Eang), \\
    \Eang' &=& \Eang, \\
    \Sb{\zb}'&=& -\Sb{\zb} +2n_3(\hat{\rr},\Eang)[\nnb\cdot\SSb],\\
    \Jb{\zb}'&=& \Jb{\zb}+2\Sb{\zb}-2n_3(\hat{\rr},\Eang)[\nnb\cdot\SSb],\\
    \Sb{b}' &=& \Sb{\zb}, \\ 
    \Jb{b}' &=& \Jb{b}+\Sb{b}-\Sb{\zb}\label{eq:Jb'},
\eea
where we define the operators $\Jb{b}=\BB(\hat{\rr},\Eang)\cdot\JJop/\vert\BB(\hat{\rr})\vert$ and $\Sb{b}=\BB(\hat{\rr},\Eang)\cdot\SSb/\vert\BB(\hat{\rr})\vert$, which represent the projection along the local direction of the magnetic field. Note, for instance, that the mathematical operator $\Sb{\zb}$ in the transformed frame has the physical meaning of the projection of the spin along the local magnetic field, as expected. However this is not the case for $\Jb{\zb}$.

In the following sections some assumptions are considered in the state of the nanomagnet to perform several approximation to the very rich, but complicated, Hamiltonian $\Hop'$ given in \eqnref{eq:Unitary_Ham_def}.

%%%%%%%%%%%%%%%%%%%%%%%%%%%%%%%%%%%%%%%%%%%%%%%%%%%%%%%%%%%%%%%%%
\section{Lamb-Dicke Approximation}\label{sec:Lamb-Dicke}
%%%%%%%%%%%%%%%%%%%%%%%%%%%%%%%%%%%%%%%%%%%%%%%%%%%%%%%%%%%%%%%%%

The first assumption is to consider that  the state of the nanomagnet is such that $\avg{\Sb{\zb}} >0 $. This implies that the term $\gamma\hbar\vert\BB(\hat \rr)\vert\Sb{\zb}$ in $\Hop_0$, see \eqnref{eq:H_0}, can confine the nanomagnet around $\rr \approx 0$ for a magnetic field with a local minima of $\vert\BB(\rr)\vert$ at $\rr  =0$~\citep{GOV}. This allows to expand the functions of $\hat \rr$ in the Hamiltonian \eqnref{eq:Unitary_Ham_def} as a Taylor expansion. Note that only $\Hop_0$, $\Vop_D$, and $\Vop_P$ depend on $\hat \rr$. The Lamb-Dicke approximation will consist in only keeping the lower orders of such expansions.

\begin{figure}[t]
	\includegraphics[width= \columnwidth]{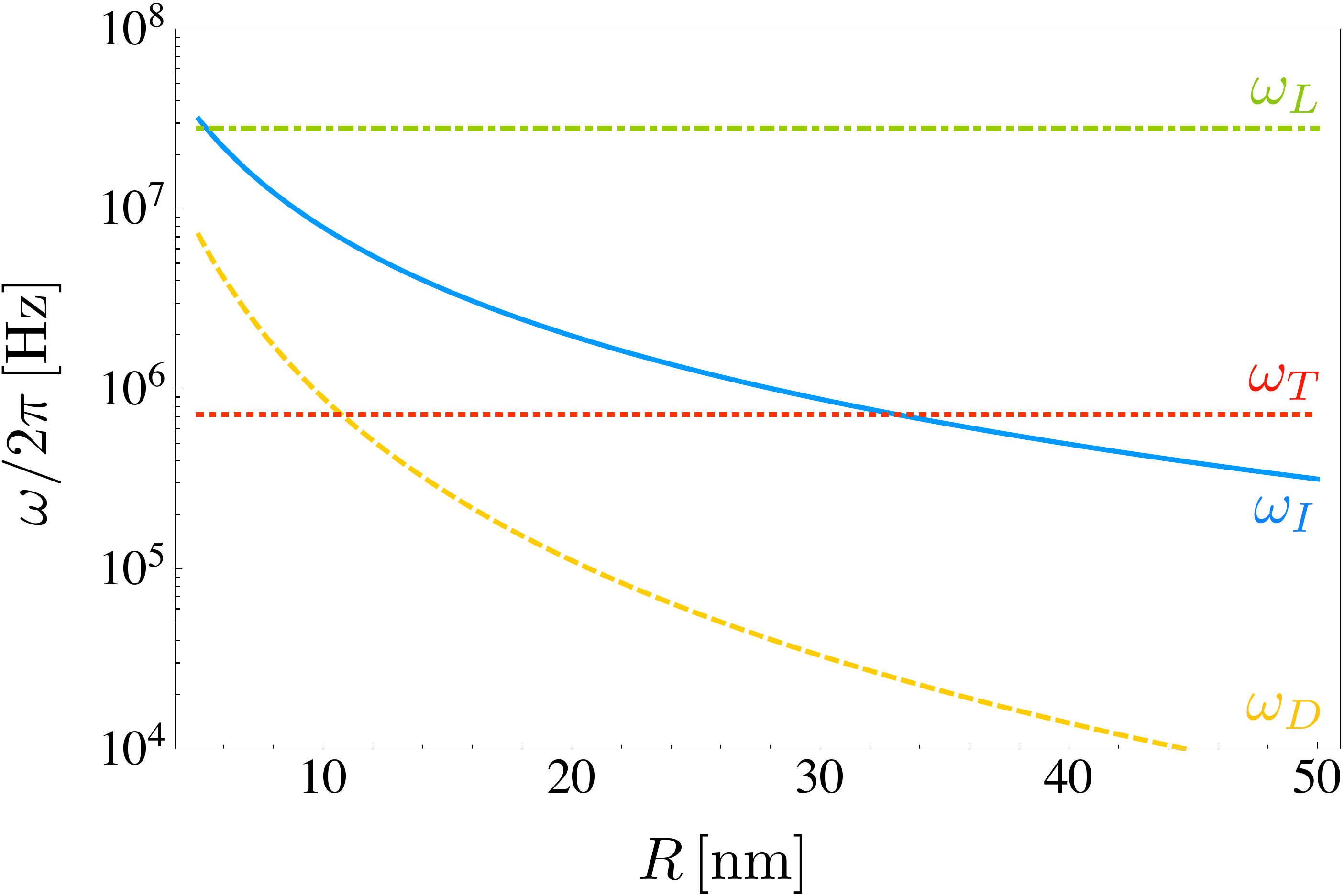}
	\caption{(Color online) The frequencies $\wB$, $\wR$, $\wA$ and $\wP$ are plotted as a function of the radius $R$ of the nanomagnet. We consider a Cobalt nanomagnet of density  $\rho_{\text{Co}}=8.9\times 10^3 \text{Kg}/m^3$, $S/N =1.7$, $T_b=\hbar^2 D\Scn^2/K_b= 30 \text{K}$. The external Ioffe-Pritchard field is used with $B_0 = 10^{-3}\text{T}$, $B' = 10^{4}\text{T}/\text{m}$ and $B'' = 10^{6}\text{T}/\text{m}^2$ \citep{PhysRevA.56.2451,PhysRevA.74.035401}.}
	\label{Fig:Plot_Freq}
\end{figure}

In $\Hop_0$, this is done by expanding
\be\label{eq:Expansion_ModulusB}
	|\BB(\hat \rr)| \simeq  B_0 + \sum_\nu\frac{\B{\nu} \hat r_\nu^2}{2},
\ee
where we have defined $B_0 \equiv |\BB(0)|$ and $\B{\nu} \equiv (\pa{\nu}^2|\BB(\rr)|)_{\rr=0}$. This expansion is valid provided
\be\label{eq:LD_Condition_B}
\abs{\frac{\avg{\hat r_\nu^2} \B{\nu}}{B_0}} \ll 1 \quad \forall \nu.
\ee
In full generality the cross talking terms coupling different spatial directions can be made zero by choosing the laboratory frame axis where the symmetric matrix $(\partial_\nu \partial_\mu |\BB(\rr)| )_{\rr=0}$ is diagonal. Under the Lamb-Dicke approximation the term $\Hop_0$ is approximated to
\be\label{eq:Ham0}
	\begin{split}
		\Hop_0 \approx & \frac{\ppop^2}{2M} + \frac{\Sb{\zb}}{\Scn} \sum_\nu\frac{M}{2} \wT{\nu}^2 \hat r_\nu^2 +\hbar\wB\Sb{\zb}- \frac{\hbar\wA}{\Scn}\Sb{\zb}^2\\
		&+\frac{\hbar\wR}{2\Scn}\pare{\JJop^2 +2\Jb{\zb}\Sb{\zb}},
	\end{split}
\ee
where we have defined the following frequencies
\bea
\wT{\nu} &\equiv& \sqrt{\frac{\hbar\gamma \B{\nu}}{M_s}},\\
\wR&\equiv& \frac{\hbar\Scn}{I}, \\
\wA&\equiv& \hbar D \Scn, \\
\wB &\equiv& \gamma B_0,
\eea
with $M_s \equiv M/\Scn$. Typical values and hierarchy of these frequencies depend on the size of the particle and on the external bias field $B_0$, see \figref{Fig:Plot_Freq}. The center-of-mass position operator can be expressed as
\be
	\hat r_\nu = \frac{\zpm{\nu}}{\sqrt{\Scn}}\pare{\cop{\nu}+\cdop{\nu}},
\ee
where $\zpm{\nu} = [\hbar/(2M_s\wT{\nu} )]^{1/2}$ and $[\cop{\nu},\cdop{\mu}]=\delta_{\nu \mu}$. 

The $\hat \rr$-dependence of the terms $\Vop_D$ and $\Vop_P$ arises from the components of $\nn(\hat \rr)$. Under the Lamb-Dicke approximation one can use the expansion
\be\label{eq:nnl_expansion}
\begin{split}
	n_{\mu}(\hat \rr) \approx &\, n_{\mu}(0)+ \sum_\nu  \frac{\zpm{\nu}}{\sqrt{\Scn}}\pare{\cop{\nu}+\cdop{\nu}}\pa{\nu}n_{\mu}\\
	&+\frac{1}{2}\sum_{\nu\lambda}\frac{\zpm{\nu}\zpm{\lambda}}{\Scn}\pare{\cdop{\nu}+\cop{\nu}}\pare{\cdop{\lambda}+\cop{\lambda}}\pa{\nu}\pa{\lambda}n_{\mu},
\end{split}
\ee
where we define $\pa{\nu}n_{\mu} = \pa{\nu} n_{\mu}(\rr) |_{\rr=0}$ for shortness.
Since the vector $\nn(\rr)$ defined in \secref{sec:Diag_MagneticDipole} depends on $\rr$ only through the magnetic field $\BB(\rr)$, this expansion is justified when the condition \eqnref{eq:LD_Condition_B} is fulfilled.
Recall that to obtain the Taylor expansion for the components of $\nnb$, it is sufficient to use \eqnref{eq:nnl2nnb} and \eqnref{eq:nnl_expansion}. This allows to expand  $\Vop_D$ and $\Vop_P$ in powers of $\hat \rr$, see \appref{apdx:Calculation_V_D} \& \ref{apdx:Calculation_V_P}. 

%%%%%%%%%%%%%%%%%%%%%%%%%%%%%%%%%%%%%%%%%%%%%%%%%%%%%%%%%%%%%%%%%
\section{Holstein-Primakoff approximation: Bosonization}\label{sec:Bosonization_AM}
%%%%%%%%%%%%%%%%%%%%%%%%%%%%%%%%%%%%%%%%%%%%%%%%%%%%%%%%%%%%%%%%%

Our aim is to describe the nanomagnet in the following regime
\bea
\avg{\hat \rr}&\approx& 0, \label{eq:avg_r_condition}\\
\avg{\Sb{\zb}} &\lesssim& S, \\
\avg{\JJop^2} &\approx& \avg{\SSb^2} = S(S+1), \\
\avg{\Jb{\zb}} &\approx& \avg{\Jl{\zl}} \gtrsim - J \approx - S\label{eq:avg_Jbz_conditon}.
\eea
Recall that for a nanomagnet $S \gg 1$. We called this regime {\em highly polarized}, which corresponds to the nanomagnet being:
\begin{itemize} 
\item Highly confined in position.
\item With the macrospin highly anti-aligned to the external magnetic field (recall that $\SSb= - \SSl$).
\item Nearly not rotating, namely $\avg{\LLop^2} \gtrsim 0$.
\end{itemize}
According to \eqnref{eq:Jb'}, we can approximate $\Jb{b}' \approx \Jb{\zb}$ when the nanomagnet is nearly not spinning.

We apply a Holstein-Primakoff (HP) boson mapping \cite{PhysRev.58.1098} to express the angular momentum operators and the D-matrices as a function of a set of bosonic operators. This mapping allows us to express $\Hop'$, \eqnref{eq:Unitary_Ham_def}, in terms of bosonic operators. The exact mapping yields non-linear bosonic terms to keep the angular momentum character of the original operators. Nevertheless, by assuming the highly polarized regime of the nanomagnet, a quadratic Hamiltonian in bosonic operators can be obtained under the so-called HP approximation.

\subsection{Holstein-Primakoff boson mapping}

The HP boson mapping for the spin angular momentum is given by \citep{PhysRev.58.1098}
\be\label{eq:ExactHP_S}
\begin{split}
	\Sb{\zb} &= \Scn -\hat s^\dag\hat s,\\
	\Sbp&= \Sbm^\dag=\pare{2\Scn-\hat s^\dag\hat s}^{1/2} \hat s .
\end{split}
\ee
We have introduced the spin bosonic operator $[\hat s, \hat s^\dag]=1$.  In the highly polarized regime, one can perform the HP approximation, namely
\be\label{eq:HP_S}
\begin{split}
	\Sb{\zb} & = \Scn - \hat s^\dag \hat s, \\
	\Sbp & = \Sbm^\dag \approx \sqrt{2\Scn}\hat s.
\end{split}
\ee

The boson mapping  for the angular momentum operator $\JJop$ and the D-matrices has to be done more carefully since $\Jcn$ is not fixed, as discussed in~\eqnref{eq:Action_Dmatrix}. The HP map can be generalized by promoting the quantum number $\Jcn$ to the operator $\Jb{} \equiv \ddop\dop/2$ with $[\dop,\ddop]=1$~\cite{RevModPhys.63.375}. The exact mapping  is  given by 
\be\label{eq:ExactHP_J}
\begin{split}
	\Jb{\zb} &= -\frac{\ddop\dop}{2} + \hat k^\dag\hat k, \\
	 \Jbp & = \Jbm^\dag=  \hat k^\dag\pare{\ddop\dop-\hat k^\dag\hat k}^{1/2}, \\
	\Jl{\zl} &= -\frac{\ddop\dop}{2} + \hat m^\dag\hat m, \\
	 \Jlp &= \Jlm^\dag =\hat m^\dag \pare{\ddop\dop-\hat m^\dag\hat m}^{1/2},\\
	 \JJop^2 &= \frac{\ddop\dop}{2}\pare{\frac{\ddop\dop}{2}+1}.
\end{split}
\ee
We have introduced the angular momentum bosonic operators $[\hat m,\hat m^\dag]=[\hat k,\hat k^\dag]=1$. One can show that the mapping fulfills the commutation rules given in \tabref{TAB:commut}. Therefore, the general state $\ket{\Jcn \M{\Jcn} \K{\Jcn}, \Scn \K{\Scn}}_2 $ can be obtained, using the bosonic mapping, from the vacuum of $4$ bosonic modes, namely
\be \label{eq:AngulartoBosons}
\begin{split}
&\ket{\Jcn \M{\Jcn} \K{\Jcn}, \Scn \K{\Scn}}_2 = \\& = \mathcal{N} \pare{\ddop}^{2J} \pare{\hat m^\dag}^{\Jcn+\M{\Jcn}} \pare{\hat k^\dag}^{\Jcn+\K{\Jcn}} \pare{\hat s^\dag}^{\K{\Scn}} \ket{0},
\end{split}
\ee
where $\mathcal{N} = [(2\Jcn)!(\Jcn+\M{\Jcn})!(\Jcn+\K{\Jcn})!\K{\Scn}!]^{-1/2}$. In the highly polarized regime of the nanomagnet, it is useful to define $\dop \equiv \sqrt{2\Jcn}+\jop$, where $\Jcn\gg 1$. In this case the HP approximation of \eqnref{eq:ExactHP_J} reads 
\be\label{eq:HP_J}
\begin{split}
	\Jb{\zb} &= -\Jcn -\frac{\sqrt{2\Jcn}}{2}\pare{\jdop+\jop} -\frac{\jdop\jop}{2} + \kdop \kop, \\
	 \Jbp & = \Jbm^\dag \approx \sqrt{2\Jcn} \kdop + \frac{\jdop+\jop}{2}\kdop + O\pare{\frac{1}{\sqrt{\Jcn}}}, \\
	\Jl{\zl} &= -\Jcn -\frac{\sqrt{2\Jcn}}{2}\pare{\jdop+\jop} -\frac{\jdop\jop}{2} + \mdop \mop, \\
	 \Jlp &= \Jlm^\dag \approx \sqrt{2\Jcn} \mdop + \frac{\jdop+\jop}{2}\mdop + O\pare{\frac{1}{\sqrt{\Jcn}}}, \\
	\end{split}
\ee
and for the total angular momentum $\JJop^2$
\be\label{eq:HP_Jtot}
\begin{split}
	\JJop^2 =& \Jcn\pare{\Jcn+1}+\Jcn\sqrt{2\Jcn}\pare{\jdop+\jop} \\ 
	&+ \Jcn\spare{\frac{(\jop+\jdop)^2}{2}+\jdop\jop}\\
	&+\frac{\sqrt{2\Jcn}}{2}\spare{\frac{\acom{\jdop+\jop}{\jdop\jop}}{2}+\pare{\jdop+\jop}}\\
	&+\frac{\jdop\jop}{2}\pare{\frac{\jdop\jop}{2}+1}.
\end{split}
\ee

Let us now address the boson mapping of the D-matrices~\citep{PhysRevC.11.1426}. This can be done using \eqnref{eq:Action_Dmatrix} and \eqnref{eq:AngulartoBosons}. Knowing how the single creation and annihilation bosonic operators act on $\ket{\Jcn \M{\Jcn} \K{\Jcn}, \Scn \K{\Scn}}_2 $, we can derive an expression of $\D{mk}{q}$ in terms of $\{\hat j, \hat j^\dag, \hat m, \hat m^\dag, \hat k, \hat k^\dag \}$, such that when applied to $\ket{\Jcn \M{\Jcn} \K{\Jcn}, \Scn \K{\Scn}}_2 $ using the mapping \eqnref{eq:AngulartoBosons} gives exactly the right hand side of \eqnref{eq:Action_Dmatrix}. While this can be done exact, in \tabref{TAB:BosonicExpr_Dmatrix} we provide the mapping, using the HP approximation up to order $O(1/\Jcn)$, for the $\D{mk}{q} $ appearing in the Hamiltonian.

\begin{table*}[t]
\caption{Bosonized expression, up to order $O\pare{1/\Jcn}$, for the relevant $\D{mk}{j}$ appearing in the Hamiltonian.}\label{TAB:BosonicExpr_Dmatrix}
\begin{ruledtabular}
\begin{tabular}{l}
\\
$j=1$:\\
\\
$\D{00}{1} = 1-\frac{1}{\Jcn}\spare{1+\kdop\kop+\mdop\mop-\kdop\mdop-\kop\mop} + O\pare{\frac{1}{\Jcn\sqrt{\Jcn}}}$\\
$\D{01}{1}=-\spare{\D{0-1}{1}}^\dagger=\frac{1}{\sqrt{\Jcn}}\pare{\kdop-\mop} -\frac{1}{2\sqrt{2}}\frac{1}{\Jcn}\spare{\pare{\jdop+\jop}\kdop+\pare{\jop-3\jdop}\mop} + O\pare{\frac{1}{\Jcn\sqrt{\Jcn}}}$\\
$\D{10}{1}=-\spare{\D{-10}{1}}^\dagger=\frac{1}{\sqrt{\Jcn}}\pare{\mdop-\kop} -\frac{1}{2\sqrt{2}}\frac{1}{\Jcn}\spare{\pare{\jdop+\jop}\mdop+\pare{\jop-3\jdop}\kop}+O\pare{\frac{1}{\Jcn\sqrt{\Jcn}}}$\\
$\D{11}{1}=\spare{\D{-1-1}{1}}^\dag=1-\frac{\jdop-\jop}{\sqrt{2\Jcn}}-\frac{1}{2\Jcn}\spare{2\pare{\jdop\jop+1}+\kdop\kop + \mdop\mop-2\mdop\kdop+\jop^2-\frac{3}{4}\pare{\jdop+\jop}^2-\frac{1}{4}\pare{\jdop-\jop}^2} + O\pare{\frac{1}{\Jcn\sqrt{\Jcn}}}$\\
$\D{1-1}{1}=\spare{\D{-11}{1}}^\dag = \frac{1}{2\Jcn}\pare{\kop^2-2\mdop\kop+(\mdop)^2}+O\pare{\frac{1}{\Jcn\sqrt{\Jcn}}}$\\
\\
$j=2$:\\
\\
$\D{00}{2} = 1-\frac{3}{\Jcn}\spare{1+\kdop\kop+\mdop\mop-\kdop\mdop-\kop\mop} + O\pare{\frac{1}{\Jcn\sqrt{\Jcn}}}$\\
$\D{01}{2}=-\spare{\D{0-1}{2}}^\dagger=\sqrt{\frac{3}{\Jcn}}\pare{\kdop-\mop} -\frac{\sqrt{3}}{2\sqrt{2}}\frac{1}{\Jcn}\spare{\pare{\jdop+\jop}\kdop+\pare{\jop-3\jdop}\mop} + O\pare{\frac{1}{\Jcn\sqrt{\Jcn}}}$\\
$\D{10}{2}=-\spare{\D{-10}{2}}^\dagger= \sqrt{\frac{3}{\Jcn}}\pare{\mdop-\kop} -\frac{\sqrt{3}}{2\sqrt{2}}\frac{1}{\Jcn}\spare{\pare{\jdop+\jop}\mdop+\pare{\jop-3\jdop}\kop}+O\pare{\frac{1}{\Jcn\sqrt{\Jcn}}}$\\
\\
$j=4$:\\
\\
$\D{00}{4} = 1-\frac{10}{\Jcn}\spare{1+\kdop\kop+\mdop\mop-\kdop\mdop-\kop\mop} + O\pare{\frac{1}{\Jcn\sqrt{\Jcn}}}$\\
$\D{01}{4}=-\spare{\D{0-1}{4}}^\dagger=\sqrt{\frac{10}{\Jcn}}\pare{\kdop-\mop} -\frac{\sqrt{10}}{2\sqrt{2}}\frac{1}{\Jcn}\spare{\pare{\jdop+\jop}\kdop+\pare{\jop-3\jdop}\mop} + O\pare{\frac{1}{\Jcn\sqrt{\Jcn}}}$\\
$\D{10}{4} =-\spare{\D{-10}{4}}^\dagger=\sqrt{\frac{10}{\Jcn}}\pare{\mdop-\kop} -\frac{\sqrt{10}}{2\sqrt{2}}\frac{1}{\Jcn}\spare{\pare{\jdop+\jop}\mdop+\pare{\jop-3\jdop}\kop}+O\pare{\frac{1}{\Jcn\sqrt{\Jcn}}}$\\
\\
\end{tabular}
\end{ruledtabular}
\end{table*}

\section{Quadratic bosonized Hamiltonian} \label{sec:quadratic}

At this stage one can use the Lamb-Dicke approximation and the HP mapping of the angular momentum operators and the D-matrices to write the Hamiltonian $\Hop'$, \eqnref{eq:Unitary_Ham_def}, as a function of the bosonic modes of the system: $\{ \cop{\xl}, \cop{\yl}, \cop{\zl}, \hat j, \hat s, \hat k, \hat m\}$ . In principle this Hamiltonian can be divided into two parts:
\be\label{eq:Ham_G+NG}
\Hop' = \hat H_{G} + \hat H_{NG}.
\ee
$\hat H_G$ collects  terms containing up to two bosonic operators (quadratic form), which lead to Gaussian physics. $\hat H_{NG}$ contains normally ordered terms with more than two bosonic operators. Let us first obtain $\hat H_G$ and then discuss when $\hat H_{NG}$ can be neglected.

\subsection{Quadratic terms}\label{sec:Gss}

In order to obtain $\Hop_G$ one substitutes \eqnref{eq:HP_S}, Eq.~(\ref{eq:HP_J}-\ref{eq:HP_Jtot}), and \tabref{TAB:BosonicExpr_Dmatrix} into the Hamiltonian \eqnref{eq:Unitary_Ham_def} after doing the Lamb-Dicke approximation. After normal-ordering  the bosonic operators one obtains the quadratic Hamiltonian of the system. 
The quadratic terms arising from $\Hop_0$, \eqnref{eq:Ham0}, and $\Hop_I$, \eqnref{eq:V_I}, are straightforward to obtain.
The bosonization of the terms $\Vop_D$, \eqnref{eq:V_D},  and $\Vop_P$, \eqnref{eq:V_P}, is more involved, see details in \appref{apdx:Calculation_V_D} \& \ref{apdx:Calculation_V_P}. 

In deriving the quadratic Hamiltonian, we assume the Ioffe-Pritchard configuration for the external magnetic field. This magnetic field is characterized by three parameters: the bias field $B_0$, the field gradient along $z$, $B'$, and the field curvature along $z$, $B''$, see \appref{apdx:Ioffe-Pritchard_Field} for details. With this field one has that $\w_\xl=\w_\yl\equiv \wP$, and the first order terms in the Lamb-Dicke expansion of $\nnb$ can be characterized by a single expansion parameter $\LD=\zpm{}B'/(2B_0)$, where $\zpm{}\equiv\zpm{x}=\zpm{y}$.

After some effort one obtains the following quadratic Hamiltonian:
\be\label{eq:Gss_Ham}
\Hop_{G}= \Hop_{0} +\hat V_s + \hat V_b,
\ee
where
\begin{itemize}

\item $\Hop_0$ is given by:
\be\label{eq:GssHam_Diag}
\begin{split}
	\frac{\Hop_{0}}{\hbar} = &\w_r \crdop\crop +\w_l \cldop\clop +\wT{\zl}\cdop{\zl}\cop{\zl} -\Delta \hat s^\dag\hat s\\
	&+\w_j \jdop\jop +\xi_1 \pare{\jdop+\jop}+\xi_{2} \pare{\jdop+\jop}^2 \\
	& + \w_{m}\hat m^\dag\hat m +\w_{k}\hat k^\dag\hat k \\& - \w_{m}\pare{\hat m^\dag \hat k^\dag+\hat m \hat k}.
\end{split}
\ee
The frequencies are given in \tabref{TAB:Freq_Definition}. We define the modes
$\hat b_{r(l)} =(\cop{\xl} \mp \im \cop{\yl})/\sqrt{2}$ describing a counter clock wise (clock wise) circular motion of the center of mass. Note that $\Delta$ can be controlled via the strength of $B_0$ and can, in principle, be either positive or negative (see \figref{Fig:Plot_Freq}). As seen below the mode $\jop$ is decoupled from the other modes in the quadratic Hamiltonian.

\item $\hat V_s$ is given by:
\be\label{eq:GssHam_JS}
\begin{split}
	\frac{\hat V_s}{\hbar} = & g_k \pare{ \hat s^\dag\hat k^\dag +  \hat s \hat k  } -g_m \pare{\hat s^\dag\hat m + \hat s \hat m^\dag}
\end{split}
\ee
This term describes the coupling between the spin degree of freedom and the total angular momentum. The coupling strengths are given in \tabref{TAB:Freq_Definition}. In essence, this term describes the Einstein-de Haas effect. 

\item $\hat V_b$ is given by:
\be\label{eq:GssHam_GR}
\begin{split}
	\frac{\hat V_b}{\hbar} =& g_l \clop\pare{\hat m^\dag-\hat k}+ g_r \crdop\pare{\mdop-\kop }\\
	& -g_s \sdop \pare{\crdop-\clop}+ \hc
\end{split}
\ee
This term describes the coupling between the center-of-mass motion and the angular momenta. The coupling strengths are given in \tabref{TAB:Freq_Definition}. Note that the coupling is induced due to the non-zero magnetic field gradient since $g_r=g_l=g_s=0$ for $B'=0$. 

\end{itemize}

\begin{table}[t!]
\caption{Definition of the frequencies appearing in \cref{eq:GssHam_Diag,eq:GssHam_JS,eq:GssHam_GR} in terms of the fundamental frequencies of the system, $\wR,\wB,\wA$ and $\wT{\nu}$. In the expression below we neglected terms smaller than $O\pare{\w \eta}$, where $\w=\wR,\wB,\wA$ or $\wT{\nu}$.}\label{TAB:Freq_Definition}
\begin{ruledtabular}
\begin{tabular}{ll}
$\w_r \crdop\crop $ & $\w_r \equiv \wP$\\
$\w_l \cldop\clop $ & $\w_l\equiv\wP$\\
$-\Delta \hat s^\dag\hat s $ & $\Delta \equiv \wB -2\wA -\wR \frac{\Jcn}{\Scn}$\\
$\w_{k}\hat k^\dag\hat k $ & $\w_{k} \equiv \wR +8\wA\frac{\Scn}{\Jcn}$\\
$\w_{m}\mdop\hat m $ & $\w_{m} \equiv 8\wA \frac{\Scn}{\Jcn}$\\
$\w_{j}\jdop\jop$ & $\w_{j} = \frac{\wR}{2}\pare{\frac{\Jcn}{\Scn}-1}$\\
$\xi_1 \pare{\jdop+\jop} $ & $\xi_1= \frac{\sqrt{2 \Jcn}}{2}\wR \pare{\frac{\Jcn}{\Scn}-1}$\\
$\xi_2\pare{\jdop+\jop}^2$ & $\xi_{2} = \frac{\wR}{4}\frac{\Jcn}{\Scn}$\\
$g_k\hat s \hat k $ & $g_k \equiv \wR\sqrt{\frac{\Jcn}{\Scn}} +4\wA\sqrt{\frac{\Scn}{\Jcn}}$\\
$g_m\hat s \hat m^\dag$ & $ g_m \equiv 4\wA\sqrt{\frac{\Scn}{\Jcn}}$\\
$g_r \crdop\pare{\mdop-\kop } $ & $g_r \equiv \LD\pare{ 8\wA+2\wP}\sqrt{\frac{\Scn}{\Jcn}}$\\
$g_l \clop\pare{\hat m^\dag-\hat k}$ & $g_l \equiv \LD\pare{8\wA-2\wP}\sqrt{\frac{\Scn}{\Jcn}}$\\
$g_s \sdop \pare{\crdop-\clop}$ & $g_s\equiv 2\wP\LD$\\
\end{tabular}
\end{ruledtabular}
\end{table}

The derivation of the quadratic Hamiltonian $\hat H_G$ is the main result of this article. The non-quadratic contributions given by $\hat H_{NG}$ can be typically neglected, see below. $\hat H_G$ describes the quantum dynamics of the nanomagnet in the magnetic trap provided it is in the highly-polarized regime defined by Eqs.~(\ref{eq:avg_r_condition}-\ref{eq:avg_Jbz_conditon}).  Under which physical parameters the highly-polarized regime is stable, the rich physics and wealth of applications that $\hat H_G$ can describe, and how to implement it in an experimentally feasible scenario, are very interesting questions that lie beyond the scope of this article and will be addressed elsewhere~\citep{CosimoInPreparation}.

\subsection{Non-quadratic terms: when can they be neglected?}

The non-quadratic Hamiltonian $\Hop_{NG}$ collects the terms with more than two normally ordered bosonic operators $\{ \cop{\xl}, \cop{\yl}, \cop{\zl}, \hat j, \hat s, \hat k, \hat m\}$. In the highly polarized regime Eq.~(\ref{eq:avg_r_condition}-\ref{eq:avg_Jbz_conditon}) $\Jcn\approx \Scn \gg 1$, one can realize that terms with an additional bosonic operator are a factor $1/\sqrt{S}$ smaller.
All the energy terms that appear in the Hamiltonian $\Hop'=\Hop_G + \Hop_{NG}$ can be expressed in terms of the frequencies $\wP$, $\w_{\zl}$, $\wA$, $\wR$ and $\wB$. Establishing a hierarchy between these frequencies is needed to understand when the non-linear terms can be neglected. The strength of these frequencies depends on the size of the nanomagnet and on the external magnetic field. As shown in \figref{Fig:Plot_Freq} different regimes can be achieved but we consider the following one
\be\label{eq:Hierarchy_Freq}
	\wB \geq \wR \gg \wA\sim\wP,\w_{\zl}.
\ee
In order to neglect the non-quadratic contribution in the bosonized Hamiltonian, one needs to compare the order of the strongest of the non-quadratic terms in $\Hop_{NG}$ with the weakest contribution kept in $\Hop_G$ taking into account \eqnref{eq:Hierarchy_Freq}. The smallest terms in $\Hop_G$ are of the order $O\pare{\wP\LD}$ and $O\pare{\wA\LD}$. The strongest non-linear term contribution comes from the bosonization of $\Hop_I$ and reads
\be
	\Vop_{jkm} = \frac{\hbar\wR}{2\sqrt{2\Scn}}\pare{\jdop+\jop}\pare{\kdop\sdop+\kop\sop} \sim O\pare{\frac{\wR}{\sqrt{\Scn}}}.
\ee
Therefore $\Hop_{NG}$ can be neglected provided
\be
	\LD \gg \text{min}\cpare{\frac{\wR}{\wA\sqrt{\Scn}},\frac{\wR}{\wP\sqrt{\Scn}}},
\ee
which can be easily satisfied under typical experimentally feasible parameters~\citep{CosimoInPreparation}.

%%%%%%%%%%%%%%%%%%%%%%%%%%%%%%%%%%%%%%%%%%%%%%%%%%%%%%%%%%%%%%%%%
\section{Conclusion and Outlook}\label{sec:Conclusions}
%%%%%%%%%%%%%%%%%%%%%%%%%%%%%%%%%%%%%%%%%%%%%%%%%%%%%%%%%%%%%%%%%

In conclusion, we have shown how to describe the quantum dynamics of a nanomagnet with uniaxial anisotropy in a magnetic trap. Starting from the basic Hamiltonian under the macroscopin approximation, we have performed a:
\begin{enumerate}
\item Change of variables to describe the dynamics in the body frame.
\item Unitary transformation to diagonalize the magnetic dipole coupling.
\item Lamb-Dicke expansion around the trapping position of the nanomagnet.
\item Bosonic Mapping of the angular momentum operators.
\item Holstein-Primakoff expansion when the macrospin is well anti-aligned and the nanomagnet is nearly non-rotating (highly polarized regime).
\end{enumerate}
This allows to derive a quadratic Hamiltonian depending on bosonic operators that describes both the Einstein-de Haas coupling between the macrospin and the rotational angular momentum, and the magnetic-trap-induced-coupling between the center-of-mass motion and the rotational angular momentum. The non-quadratic terms can be neglected in the highly-polarized regime with typical experimental parameters.

The derivation of the Hamiltonian can be straightforwardly extended to cover more general situations. For instance, one could consider ellipsoidal objects instead of spherical and other forms of magnetic anisotropy (\eg~cubic anisotropy). This  would change the rotational kinetic and the anisotropy magnetic term, respectively, in the very initial Hamiltonian but the derivation to obtain the quadratic Hamiltonian would be analogous. Should one also be interested in describing the magnetic coupling of the nanomagnet with the quantum electromagnetic field, one could just add the quantized magnetic field operator to the magnetic dipole interaction term in the very initial Hamiltonian. This could be used to calculate vacuum forces and decay rates of the eigenstates of the Hamiltonian. Also note that the derivation of the quadratic Hamiltonian is constructive in the sense that one could consider next orders in both the Lamb-Dicke and the Holstein-Primakoff expansion to obtain the higher order non-quadratic contributions. 

The unitary transformation in \secref{sec:Diag_MagneticDipole} is motivated by the goal of making the magnetic trapping potential appear explicitly in the Hamiltonian of the system. However, this transformation is not necessary to describe the dynamics of the quantum fluctuations around a stable state of the system. One could carry out the bosonization of the fluctuations in the Hamiltonian \eqnref{eq:Ham_Body}, before applying the unitary transformation. While this is mathematically equivalent, one should nevertheless be careful when performing the approximations after bosonizing since the physical meaning of the fluctuations, should one bosonized the same mathematical operators, would be different in the two cases. Recall the  remark made in the end of \secref{sec:Diag_MagneticDipole}. 
%The unitary transformation in \secref{sec:Diag_MagneticDipole} is motivated by the goal of making the harmonic trapping potential to appear explicitly in the Hamiltonian of the system. This transformation is not necessary in describing the dynamics of the quantum fluctuation of the system around its equilibrium positions. We could have in principle carried out the linearization of the degrees of freedom around the equilibrium position of the system starting from the Hamiltonian \eqnref{eq:Ham_Body}. Since a unitary transformation leaves the dynamic invariant the two systems are equivalents, however the meaning of the fluctuations is different. In the case studied here the quantum fluctuations of the angular momentum are referred to the local direction of the magnetic field, while in this latter approach the fluctuation would be referred to the anisotropy axis. Because of the approximation introduced in \secref{sec:quadratic} these two point of view could lead to essential differences in the dynamics of the linearised system, while leaving the dynamic of the exact nonlinear system, \eqnref{eq:Ham_Body}, unchanged.

The derived quadratic Hamiltonian can be handled with the usual techniques of quantum optics and describes a very rich and original physical system. This opens many further directions that we are currently investigating~\cite{CosimoInPreparation}. Some of them are: (i) an experimental proposal where nanomagnets can be levitated in a $Z$-shaped atom chip and brought to the quantum regime by inductively coupling its motion to a flux-dependent quantum circuit. This coupling can be used to cool the center-of-mass motion to the ground state as well as to sympathetically cool the rotational angular momentum and the macroscopin degrees of freedom;  (ii) The proposal and study of magnetically levitating the nanomagnet using the Meissner field induced on a superconductive film with a hole~\citep{ColinT}. This permanent passive short-distance trap could be used to measure Casimir and short-distance gravitational forces; (iii) Once in the quantum regime, exploit the nanomagnet as a nano-gyroscope; (iv) Perform matter-wave interferometry by releasing the ground-state cooled nanomagnet from the trap; (v) Explore the regimes where the non-Gaussian physics given by the non-quadratic terms are relevant, for instance in the context of superradiance effects, etc.

We acknowledge discussions with J. I. Cirac. We thank V. P\"ochhacker for carefully reading the manuscript. This work is supported by the European Research Council (ERC-2013-StG 335489 QSuperMag) and the Austrian Federal Ministry of Science, Research, and Economy (BMWFW).

\appendix

%%%%%%%%%%%%%%%%%%%%%%%%%%%%%%%%%%%%%%%%%%%%%%%%%%%%%%%%%%%%%%%%%
\section{Properties of the Wigner D-matrix tensor operator}\label{apdx:Wigner_Dmatrix}
%%%%%%%%%%%%%%%%%%%%%%%%%%%%%%%%%%%%%%%%%%%%%%%%%%%%%%%%%%%%%%%%%

In this section we list some useful properties of the D-matrix tensor operator. We will not prove them. The interested reader is referred to \citep{edmonds1996angular,biedenharn1981angular} for more details. These are:
\begin{enumerate}
\item The matrix elements of $\D{mk}{q}$ between the eigenstates $\ket{\Lcn\M{\Lcn}\K{\Lcn}}$ of the operators $\LLop^2,\Ll{\zl}$ and $\Lb{\zb}$ read
\be\label{eq:D_elem_Lbasis}
\begin{split}
    &\bra{\Lcn'\M{\Lcn}'\K{\Lcn}'}\D{mk}{q}\ket{\Lcn\M{\Lcn}\K{\Lcn}}=\\
    &=\frac{8\pi^2}{2\Lcn'+1}\CG{qm,\Lcn\M{\Lcn}}{\Lcn'\M{\Lcn}'}\CG{qk,\Lcn\K{\Lcn}}{\Lcn'\K{\Lcn}'}.
\end{split}
\ee

\item  The matrix elements of $\D{mk}{q}$ on the basis $\ket{\Jcn\M{\Jcn}\K{\Jcn},\Scn\K{\Scn}}_2$ read
\be\label{eq:D_matrix_element}
\begin{split}
	&{}_2\bra{\Jcn'\M{\Jcn}'\K{\Jcn}',\Scn'\K{\Scn}'}\D{mk}{q}\ket{\Jcn\M{\Jcn}\K{\Jcn},\Scn\K{\Scn}}_2= \\
	& = \delta_{\Scn\Scn'}\delta_{\K{\Scn}\K{\Scn}'} \sqrt{\frac{2\Jcn+1}{2\Jcn'+1}}\CG{qm,\Jcn\M{\Jcn}}{\Jcn'\M{\Jcn}'}\CG{qk,\Jcn\K{\Jcn}}{\Jcn'\K{\Jcn}'}
\end{split}
\ee
This result can be proved by changing the basis to $\ket{\Lcn\M{\Lcn}\K{\Lcn},\Scn\K{\Scn}}_1$ through the \eqnref{eq:H1toH2} and \eqnref{eq:H2toH3} and then applying  \eqnref{eq:D_elem_Lbasis}.\\

\item The product of two D-matrix operators reads
\be\label{eq:Identities_WignerD}
\begin{split}
	&\D{m_1k_1}{q_1}\D{m_2k_2}{q_2}=\\ &=\sum_{Q=|q_1-q_2|}^{q_1+q_2}\sum_{M,K=-Q}^{Q}\CG{q_1m_1,q_2m_2}{QM}\CG{q_1k_1,q_2k_2}{QK}\D{MK}{Q}
\end{split}
\ee

\item The D-matrix operators fulfill the following commutation relations with the total angular momentum of the system $\JJop$:
\be\label{eq:Commut_Dmatr_Jop}
\begin{split}
	&\com{\Jl{\zl}}{\D{mk}{q}}=m\D{mk}{q}\\
	&\com{\Jlpm}{\D{mk}{q}}=\sqrt{(q\mp m)(q\pm m +1)}\D{m\pm1k}{q}\\
	&\com{\Jbz}{\D{mk}{q}}=k\D{mk}{q}\\
	&\com{\Jbp}{\D{mk}{q}}=\sqrt{(q- k)(q+ k +1)}\D{mk+1}{q}\\
	&\com{\Jbm}{\D{mk}{q}}=\sqrt{(q+ k)(q- k +1)}\D{mk-1}{q}.
\end{split}
\ee

\end{enumerate}

The strategy to prove these relations is via the properties of the function $D_{mk}^{j}(\eula,\eulb,\eulc)$, defined in \eqnref{eq:WignerD}. It can be shown that $D_{mk}^{j}(\eula,\eulb,\eulc)\sim\braket{\eula,\eulb,\eulc}{\Lcn,\M{\Lcn},\K{\Lcn}}$. These eigenfuntions satisfy the following normalization condition
\be\label{eq:D_matrix_normalizz}
\begin{split}
\int_{0}^{2\pi}&\text{d}\eula \int_{0}^{\pi}\text{d}\eulb \sin{\eulb}\int_{0}^{2\pi}\text{d}\eulc \pare{D_{m_1k_1}^{q_1}(\eula,\eulb,\eulc)}^*\times\\
&\times D_{m_2k_2}^{q_2}(\eula,\eulb,\eulc) = \frac{8\pi^2}{2q_1+1}\delta_{q_2,q_1}\delta_{m_2m_1}\delta_{k_1k_2}.
\end{split}
\ee
Moreover it can be shown that the product of two such eigenfunction can be written as a linear combination of single eigenfuctions with different indexes according to the following relation
\be\label{eq:ProductWignerD}
\begin{split}
	&D_{m_1k_1}^{q_1}(\Omega) D_{m_2k_2}^{q_2}(\Omega)=\\ &=\sum_{Q=|q_1-q_2|}^{q_1+q_2}\sum_{M,K=-Q}^{Q}\CG{q_1m_1,q_2m_2}{QM}\CG{q_1k_1,q_2k_2}{QK} D_{MK}^{Q}(\Omega).
\end{split}
\ee
Combining the results in \eqnref{eq:D_matrix_normalizz} and \eqnref{eq:ProductWignerD} one can prove the following identity involving the angular integral of three eigenfunction 
\be\label{eq:3Integral}
\begin{split}
    \int_{0}^{2\pi}&\text{d}\eula \int_{0}^{\pi}\text{d}\eulb \sin{\eulb}\int_{0}^{2\pi}\text{d}\eulc \spare{D_{M'K'}^{L'}(\Omega)}^*\times\\ &\times D_{mk}^{q}(\Omega)D_{MK}^{L}(\Omega) = \frac{8\pi^2}{2L'+1}\CG{qm,LM}{L'M'}\CG{qk,LK}{L'K'}.
\end{split}
\ee
The integral in \eqnref{eq:3Integral} can be used to obtain \eqnref{eq:D_matrix_element}. See \citep{edmonds1996angular,biedenharn1981angular} for further details.

%%%%%%%%%%%%%%%%%%%%%%%%%%%%%%%%%%%%%%%%%%%%%%%%%%%%%%%%%%%%%%%%%
\section{Ioffe-Pritchard Magnetic Trap}\label{apdx:Ioffe-Pritchard_Field}
%%%%%%%%%%%%%%%%%%%%%%%%%%%%%%%%%%%%%%%%%%%%%%%%%%%%%%%%%%%%%%%%%

A magnetic field in free space must satisfy the homogeneous Maxwell's equations $\div \BB(\rr)=0$ and $\curl \BB(\rr)=0$.
If one asks additionally for axial symmetry, \ie $\BB(\rho,\varphi,z)=B_{\rho}(\rho,z)\ue{\rho}+B_z (\rho,z)\elz$, the expression for the most general magnetic field close the axis of symmetry $\elz$ and satisfying these constraints reads \citep{reichel2011atom}
\be\label{eq:Ioffe_Pritchard}
\begin{split}
	\BB(\rr) =&\,\elx \pare{B' x-\frac{B''}{2}xz}-\ely\pare{B' y+\frac{B''}{2}zy}\\
	& +\elz\spare{ B_0 + \frac{B''}{2}\pare{z^2-\frac{x^2+y^2}{2}}} ,
\end{split}
\ee
where $B_0=|\BB(0)|$, $B'\equiv \pa{z}B_{\zl}(\rr)|_{\rr=0}$, and $B''\equiv \pa{\zl}^2B_z({\rr})|_{\rr=0}$. The field in \eqnref{eq:Ioffe_Pritchard} is often called Ioffe-Pritchard.

When the particle is confined at the center of the trap, namely:
\bea \label{eq:AdiabaticCondition}
	\sqrt{\avg{\zop^2}} & \ll &\sqrt{\frac{B_0}{B''}} \\
	\sqrt{\avg{\xop^2}}\approx \sqrt{\avg{\yop^2}} & \ll & \sqrt{\frac{B_0^2}{B'^2-B_0B''}} \sim \frac{B_0}{B'},
\eea 
one can approximate
\be\label{eq:QuadraticField}
	|\BB(\rr)| \simeq B_0 +\frac{B''}{2} \zl^2 + \frac{B'^2-B_0B''/2}{2B_0}\pare{\xl^2+\yl^2}.
\ee
By substituting this expansion into the Hamiltonian $\Hop_0$ in \eqnref{eq:H_0}, one obtains a three-dimensional Harmonic potential with trapping frequencies
\be
\begin{split}
    \wT{\xl}=\wT{\yl} = \wP &\equiv \sqrt{\frac{\hbar\gamma}{M_sB_0}\pare{B'^2-\frac{1}{2}B_0B''}}\\
	\w_z &= \sqrt{\frac{\hbar\gamma B''}{M_s}},
\end{split}
\ee
We define $\zpm{}\equiv\zpm{x}=\zpm{y}$. It is useful to introduce the following Lamb-Dicke parameters that characterize the confinement of the particle in the Ioffe-Pritchard magnetic trap
\bea
    \LD &=& \frac{B'\zpm{}}{2B_0},\label{eq:LD}\\
    \dLD &=& \frac{B''\zpm{}^2}{4B_0},\label{eq:dLD}
\eea
Typically $\eta'  \ll  \eta \ll 1$.
With such magnetic field, the Taylor expansion of $\nnb$ can be obtained by realizing that :
\be\label{eq:n_components_IP_field}
\begin{split}
	&\nlz(\rr=0)=1,\\
	&\pa{\xl}\nlp |_{\rr=0} = \pa{\xl}\nlm |_{\rr=0} = \frac{B'}{2B_0},\\
	&\pa{\yl}\nlp |_{\rr=0} = -\pa{\yl}\nlm |_{\rr=0} = -\im\frac{B'}{2B_0},\\
	&\pa{\xl}^2\nlz |_{\rr=0} = \pa{\yl}^2\nlz |_{\rr=0} = -\pare{\frac{B'}{2B_0}}^2,\\
	&\pa{\xl\zl}\nlp |_{\rr=0} = \pa{\xl\zl}\nlm |_{\rr=0} = -\frac{B''}{4B_0},\\
	&\pa{\yl\zl}\nlp |_{\rr=0} = -\pa{\yl\zl}\nlm |_{\rr=0} = -\im\frac{B''}{4B_0}.
\end{split}
\ee
The other components not shown are identically zero. To calculate the operators $n_{\zb}(\hat{\rr},\Eang)$, $n_{\uparrow}(\hat{\rr},\Eang)$, $n_{\downarrow}(\hat{\rr},\Eang)$ and their derivatives we just need to substitute the results of \eqnref{eq:n_components_IP_field} into \eqnref{eq:nnl2nnb}. The terms to zero order in Lamb-Dicke, $\nbz=n_{3}(0,\Eang)$, $\nbp=n_{\uparrow}(0,\Eang)$ and $\nbm=n_{\downarrow}(0,\Eang)$, read
\be\label{eq:nb_r0}
\begin{split}
	&\nbz = \D{00}{1},\\
	&\nbp=-\sqrt{2}\D{01}{1},\\
	&\nbm=(\nbp)^{\dag}.
\end{split}
\ee
The first derivative of the components of $\nnb$ evaluated in $\rr=0$, \ie the operators $\pa{\nu}\nbz=\pa{\nu}n_{3}(\hat{\rr},\Eang)|_{\rr=0},\pa{\nu}\nbp=\pa{\nu}n_{\uparrow}(\hat{\rr},\Eang)|_{\rr=0}$ and $\pa{\nu}\nbm=\pa{\nu}n_{\downarrow}(\hat{\rr},\Eang)|_{\rr=0}$, read
\be\label{eq:LDb_x}
\begin{split}
	&\pa{\xl}\nbz = -\frac{B'}{2\sqrt{2}B_0}\pare{\D{10}{1}-\D{-10}{1}}\\
	&\pa{\xl}\nbp = -\frac{B'}{2B_0}\pare{\D{-11}{1}-\D{11}{1}}\\
	&\pa{\xl}\nbm = -\frac{B'}{2B_0}\pare{\D{1-1}{1}-\D{-1-1}{1}}
\end{split}
\ee
for $\nu=\xl$, and 
\be\label{eq:LDb_y}
\begin{split}
	&\pa{\yl}\nbz = -\im\frac{B'}{2\sqrt{2}B_0}\pare{\D{10}{1}-\D{-10}{1}}\\
	&\pa{\yl}\nbp = \im\frac{B'}{2B_0}\pare{\D{-11}{1}-\D{11}{1}}\\
	&\pa{\yl}\nbm = \im\frac{B'}{2B_0}\pare{\D{1-1}{1}-\D{-1-1}{1}},
\end{split}
\ee
for $\nu=\yl$. The operators for $\nu=\zl$ are zero due to the axial symmetry of the magnetic field.
The second derivatives of the components of $\nnb$ are obtained from \eqnref{eq:n_components_IP_field} as follow
\be\label{eq:ddnb_r0}
	\vect{\pa{\nu\mu}\nbz}{\pa{\nu\mu}\nbp}{\pa{\nu\mu}\nbm} = \TT \vect{\pa{\nu\mu}\nlz}{\pa{\nu\mu}\nlp}{\pa{\nu\mu}\nlm},
\ee
where all the derivatives are evaluated in $\rr=0$.

%%%%%%%%%%%%%%%%%%%%%%%%%%%%%%%%%%%%%%%%%%%%%%%%%%%%%%%%%%%%%%%%%
\section{Bosonization of $\Vop_D$}\label{apdx:Calculation_V_D}
%%%%%%%%%%%%%%%%%%%%%%%%%%%%%%%%%%%%%%%%%%%%%%%%%%%%%%%%%%%%%%%%%

This section provides more details on the bosonization of $\Vop_D = \Vop_D^G + \Vop_D^{NG}$ into quadratic and non-qudratic terms assuming a the Ioffe-Pritchards field \eqnref{eq:Ioffe_Pritchard}.

To obtain $\Vop_D^G$ the procedure is the following:
\begin{enumerate}
\item Apply the Lamb-Dicke expansion on the components of $\nnb$ up to second order in $\hat\rr$.
\item Substitute it into \eqnref{eq:V_D} and expand the square keeping only terms up to second order in $\hat\rr$. After this procedure the anisotropy energy reads $\Vop_D = \Vop_{D0} + \Vop_{D1} + \Vop_{D2}$ where $\Vop_{Di}$ contains terms of order $i$ in $\hat\rr$.
\item Substitute the bosonic expressions for the spin angular momentum operators, \eqnref{eq:HP_S}, and for the D-matrix, \tabref{TAB:BosonicExpr_Dmatrix}.
\item Rearrange the operators in normal order and keep the terms that contain only up to product of two bosons.
\end{enumerate}

Following this procedure one can derive all the quadratic terms that appear in $\Vop_D^G= \Vop_{D0}^G + \Vop_{D1}^G + \Vop_{D2}^G$. Note that some of this quadratic terms appear from the normal ordering of non-linear terms, and will thus lead to subleading contribution. Here we report only the quadratic terms up to order $O(\hbar\wA\LD)$:
\begin{itemize}
\item $\Vop_{D0}^G$ collects the zero order terms in the Lamb-Dicke expansion of $\nnb$. Its does not contain the center-of-mass bosons, and couples only the spin with the total angular momentum.
It reads
    \be
    \begin{split}
        \Vop_{D0}^G=&32\hbar\wA \frac{\Scn}{\Jcn}\jdop\jop -\frac{56}{5}\frac{\Scn}{\Jcn}\hbar\wA\pare{\jdop+\jop}^2\\
        &+8\hbar\wA\frac{\Scn}{\Jcn}\pare{\hat k^\dag\hat k+\hat m^\dag\hat m-\hat k^\dag\hat m^\dag-\hat k\hat m} \\
	& +4\hbar\wA\sqrt{\frac{\Scn}{\Jcn}}\spare{\hat s^\dag\pare{\hat k^\dag-\hat m}+\hc}\\
	&+O\pare{\frac{\hbar\wA}{\Jcn}}
    \end{split}
    \ee
\item $\Vop_{D1}^G$ collects the first order Lamb-Dicke expansion and couples the center of mass motion with the total angular momentum $\JJop$. It reads 
    \be
    \begin{split}
    \Vop_{D1}^G=&8\hbar\wA\LD \sqrt{\frac{\Scn}{\Jcn}}\Big[\pare{\crdop+\clop}\pare{\hat m^\dag-\hat k}+\hc\Big]\\
    & + O\pare{\frac{\hbar\wA}{\sqrt{\Scn\Jcn}}\LD}.
    \end{split}
    \ee
\item $\Vop_{D2}^G$ collects the second order expansion in Lamb-Dicke. It is a quadratic term in only the center-of-mass bosons but it is subleading, since $\Vop_{D2}^G \simeq O(\hbar \wP \LD^2)$.
\end{itemize}

The procedure to collect the non-linear contributions is analogous, but the expansions in steps 2 and 3 are carried up to the order of interest. 
Following this approach the strongest non linear term in $\Vop_D^{NG}$ is a three boson term that scales as $O(\hbar\wA \dLD/\sqrt{\Jcn})$.

%%%%%%%%%%%%%%%%%%%%%%%%%%%%%%%%%%%%%%%%%%%%%%%%%%%%%%%%%%%%%%%%%
\section{Bosonization of $\Vop_{P}$}\label{apdx:NonAdiabaticTerms}\label{apdx:Calculation_V_P}
%%%%%%%%%%%%%%%%%%%%%%%%%%%%%%%%%%%%%%%%%%%%%%%%%%%%%%%%%%%%%%%%%

This section provides more details on the bosonization of $\Vop_P = \Vop_{P1}+\Vop_{P2}$ in \cref{eq:V_P}, assuming the Ioffe-Pritchard field, where
\be\label{eq:Vop1}
	\Vop_{P1} = \frac{\AB(\hat{\rr},\Eang)\cdot\ppop + \ppop\cdot\AB(\hat{\rr},\Eang)}{2M},
\ee
is the linear term in $\AB(\hat{\rr},\Eang)$ and
\be\label{eq:Vop2}
	\Vop_{P2} = \frac{\AB^2(\hat{\rr},\Eang)}{2M},
\ee
is the quadratic term in $\AB(\hat{\rr},\Eang)$. Recall that
\be
A_{\nu}(\hat{\rr},\Eang)=-2\hbar\spare{\nnb\times\pa{\nu}\nnb}\cdot\SSb.
\ee 
We find the contribution to the final bosonic Hamiltonian $\Hop_{G}$ and estimate the strongest non-linear term in $\Hop_{NG}$. 

The procedure to obtain the quadratic terms $\Vop_P^G=\Vop_{P1}^G+\Vop_{P2}^G$, is the following:
\begin{enumerate}
\item Apply the Lamb-Dicke expansion up to second order in $\hat{\rr}$:
\be\label{eq:Expansion_A}
\begin{split}
    A_{\nu}(\hat{\rr},\Eang) \simeq & A_{\nu} + \sum_\mu\pa{\mu}A_{\nu}\hat{r}_\mu +\\
    &+\frac{1}{2}\sum_{\mu\lambda}\pa{\mu\lambda}A_{\nu}\hat{r}_{\mu} \hat{r}_{\lambda} + O(\rr^3),
\end{split}
\ee
where 
\be
\begin{split}
        A_{\nu} =& -2\hbar\pare{\nn\times\pa{\nu}\nn}\cdot \SSb\\
        \pa{\mu}A_{\nu} =& -2\hbar \spare{\pare{\nn\times\pa{\nu\mu}\nn},
	    +\pare{\pa{\mu}\nn\times\pa{\nu}\nn}}\cdot \SSb,\\
	    \pa{\mu\lambda}A_{\nu} =& -\hbar\big[\pare{\nn\times \pa{\nu\mu\lambda}\nn}+2 \pare{\pa{\mu}\nn \times \pa{\nu\lambda}\nn}+\\
	    &\pare{\pa{\mu\lambda}\nn\times\pa{\nu}\nn}\big]\cdot \SSb.
\end{split}
\ee
We define $\nn\equiv\nn(0,\Eang)$, $\pa{\nu}\nn\equiv\pa{\nu}\nn(\rr,\Eang) |_{\rr=0}$, $\pa{\nu\mu}\nn\equiv\pa{\nu\mu}\nn(\rr,\Eang) |_{\rr=0}$, $\pa{\nu\mu\lambda}\nn\equiv\pa{\nu\mu\lambda}\nn(\rr,\Eang) |_{\rr=0}$, and $\hat{r}_{\mu}=\hat{x},\hat{y},\hat{z}$ when $\mu=x,y,z$ respectively.
\item Substitute \eqnref{eq:Expansion_A} into \eqnref{eq:Vop1} and \eqnref{eq:Vop2} keeping only terms up to second order in $\hat{\rr}$.
\item Use Eq.~(\ref{eq:nb_r0}-\ref{eq:ddnb_r0}) to express $\nn, \pa{\nu}\nn, \pa{\nu\mu}\nn$ and $\pa{\nu\mu\lambda}\nn$ in terms of D-matrices, the components of $\nnl$, and its derivatives, \eqnref{eq:n_components_IP_field}.
\item Take the expression for $\Vop_{P1}$ and $\Vop_{P2}$ obtained in the preceeding step and substitute the bosonic expression for the angular momentum $\SSb$, \eqnref{eq:HP_S}, and for the D-matrix, \tabref{TAB:BosonicExpr_Dmatrix}. 
\item Rearrange the bosonic operators in normal order and keep only terms that contain product of up to two bosonic operators.
\end{enumerate}

The procedure allows to exactly calculate all the quadratic terms appearing in $\Vop_{P}^G=\Vop_{P1}^G+\Vop_{P2}^G$. Here we will report terms up to $O(\hbar \wP \LD)$. The linear terms in $\Vop_P^G$ read

\be
\begin{split}
    \Vop_{P1}^G =& 2\hbar\wP\LD\sqrt{\frac{\Scn}{\Jcn}}\spare{\pare{\mdop - \kop}\pare{\crdop - \clop} + \hc}+\\
    &+O\pare{\hbar\wP\LD^2}.
\end{split}
\ee
The contribution from the quadratic term in $\AB(\hat{\rr},\Eang)$ is of order $\Vop_{P2}^G \simeq O(\hbar\wP\LD^2)$, and therefore can be neglected.

To evaluate the non-linear terms one proceeds as for the case of the quadratic terms, however steps 1-3 must be done keeping all the bosonic terms up to the order of interest.
Following this approach one finds that the strongest non linear term in $\Vop_{P}^{NG}$ is a three bosons term that scales as $O(\hbar\wP \dLD/\sqrt{\Jcn})$.

\end{document}